\documentclass[11pt]{article}
\newcommand{\pmb}[1]{{\setbox0=\hbox{#1}%
  \kern-.025em\copy0\kern-\wd0
  \kern.05em\copy0\kern-\wd0
  \kern-.025em\raise.0433em\box0 }}
\newcommand{\fg}{\boldsymbol}
\usepackage{citesort}
\usepackage{amssymb,amsfonts,mathrsfs,color}
\usepackage{latexsym}
\usepackage{amsmath}
\usepackage{comment}
\usepackage{epstopdf}
\usepackage[pdftex]{graphicx}
\usepackage{graphicx}
\usepackage[titletoc,title]{appendix}
\numberwithin{equation}{section}
\DeclareGraphicsExtensions{.pdf,.jpg,.eps}
\usepackage{epsf}
\usepackage{subfigure}
\usepackage{calc}

\newlength{\spacer}
\newsavebox{\mybox}

\newcommand{\bey}{\begin{eqnarray}}
\newcommand{\eey}{\end{eqnarray}}

\newcommand{\bec}{\begin{center}}
\newcommand{\eec}{\end{center}}

\newcommand{\drop}[1]{}

\usepackage[lmargin=2.0cm,rmargin=2.0cm,tmargin=2.0cm,bmargin=2.0cm]{geometry}
\begin {document}
\pagestyle{myheadings}
\setcounter{tocdepth}{2}
\baselineskip22pt
\belowdisplayskip11pt
\belowdisplayshortskip11pt

\bec
{\Large A multiphase phase-field study of three-dimensional martensitic twinned  microstructures  at large strains }
\eec
\bec
{\large Anup Basak$^{1}$ and Valery I. Levitas$^{2,3}$}\\
{\it
$^{1}$ Department of Mechanical Engineering,  Indian Institute of Technology Tirupati, Tirupati, A.P. 517506, India.\\
$^2$ Departments of Aerospace Engineering, Mechanical Engineering, and  Material Science and Engineering,
 Iowa State University, Ames, IA 50011, USA.\\
 $^3$Division of Materials Science and Engineering, Ames Laboratory,   Ames, IA 50011, USA. \\
}
\eec

A thermodynamically consistent multiphase phase-field approach for stress and temperature-induced martensitic phase transformation  at the nanoscale and under large strains is developed. A total of $N$ independent order parameters are considered for materials with $N$ variants, where one of the order parameters describes  $\sf A\leftrightarrow \sf M$ transformations and the remaining $N-1$ independent order parameters describe the transformations between the variants. A non-contradictory  gradient energy is used within the free energy of the system to account for the energies of the interfaces. In addition, a non-contradictory kinetic relationships for the rate of the order parameters versus thermodynamic driving forces is suggested. As a result, a system of consistent coupled Ginzburg-Landau  equations for the order parameters are derived. The crystallographic solution for twins within twins is presented for the cubic to tetragonal transformations.  A 3D complex  twins within twins  microstructure is simulated using the developed phase-field approach and  a large-strain-based nonlinear finite element method. A comparative study between the crystallographic solution and the simulation result is presented.

\vspace{2mm}\noindent {\bf Keywords:} Multiphase phase-field approach; Martensitic transformations; Twins within twins; Crystallographic solution;  Large strains; Finite element method.

 \section{Introduction}
 \label{introd}

 {\em \bf Martensitic transformations and microstructures}. Martensitic transformations (MTs) are diffusionless solid-solid phase transformations observed in many metallic and nonmetallic crystalline solids, minerals, and various compounds, where a parent phase called austenite (high temperature phase) transforms into the product phase called martensite (low temperature phase) \cite{Bha04,Pitteri-Zanzotto-2003}. The martensitic phase (here denoted by $\sf M$) has lower crystallographic symmetry than the austenite phase (here denoted by $\sf A$), and  generally has multiple variants. Very complex microstructures such as austenite-twinned martensite, twins within twins, twins within twins within twins, wedge, X-interfaces, etc., are observed within the materials undergoing MTs \cite{Adachi-Wayman-75,Bha04,Pitteri-Zanzotto-2003,Ball-James-87,Bhattacharya-1991}. The evolution of such microstructures play a central role in, for example, strengthening of steel, shape memory effect in various alloys, ferromagnetic effect, caloric effects, etc \cite{Porter-Easterling,Bha04}.

 In continuum theories for MTs, such phase-changing materials are modeled as nonlinear elastic materials having multiple wells in the free energy density function \cite{Ball-James-87,Bha04,Pitteri-Zanzotto-2003,SteinmannCMT2014,GovindjeeCMT2007}. Fine twinned microstructures associated with austenite-martensite interfaces,  which were observed under the microscopes \cite{Schryvers-02,Chu_thesis1993,Abeyaratne-Chu_96,Schryvers-98}, are usually obtained as the minimizers of such non-convex energies within the continuum theories \cite{Ball-James-87}. The analytical crystallographic solutions for twins between a pair of variants and the austenite-twinned martensite interfaces are well-known within the small as well as finite deformation theories \cite{Ball-James-87,Bha04,Pitteri-Zanzotto-2003,James-Hane-2000}. These solutions have been further used  for more complex wedge microstructures \cite{Bhattacharya-1991,Hane-Shield1999} and X-interfaces  \cite{RuddockARM1994,StupkiewiczCMT2012}. Another important complex microstructure is twins within twins \cite{Adachi-Wayman-75,Bha04} for which the general crystallographic equations were established in \cite{Bha04,Chu_thesis1993,Abeyaratne-Chu_96}. Though the governing equations are well-known for twins within twins, the analytical solutions to such microstructures are still missing to the best of our knowledge.

 {\em \bf Phase-field approach to MTs}. The phase-field approach based on the  Ginzburg-Landau equations \cite{Umantsevbook} (similar to Allen-Cahn's approach \cite{Allen-Cahn-79}), which provides an ideal framework for studying the MTs, have been widely used for studying nucleation, growth of the phases, and evolution of complex microstructures \cite{Khachaturyan-01,Levitas-Basak-2017,Levin-Levitas-IJSS-13,Stupkiewicz-2017,Artemev-Wang-Khachaturyan-2000-Acta,Artemev-Jin-Khachaturyan-2001-Acta,Schryvers-12,Artemev-Slutsker-Roytburd-Acta,Chen-Material_sc,Chen-14,Chen-1995,Finel-2012,Steinbach-2019,SaxenaCMT2012,Rabbe-JMPS-14,PF_review,levitas-javanbakht-PRL-10,Landis-2016,levitasetal-prl-09,Levitas-Javanbakht-IJMR-11,Lei2010,Hildebrand-Miehe-12,Clayton-knap-11,Clayton2021,ClaytonCMT2014,She_2015,Levitas-preston-PRB-I,Levitas-preston-PRB-II,Levitas-IJP-13,Levitas-IJP-18,Babaei-Levitas-IJP-18}. A set of sufficiently smooth scalar internal variables called the order parameters, are used to describe the phases. The volume fraction based (e.g. \cite{Tuma-Stupkiewicz-16,Tuma-Stupkiewicz-Petryk-16,Tuma-Stupkiewicz-Petryk-18,Stupkiewicz-20,Levitas-Id-Pres-PRL-04,Idesman-Levitas-JMPS-05,Levitas-etal-PRL-18,Esfahani-Levitas-AM-20,Babaei-Levitas-JMPS-20}) or the transformation strains based (e.g. \cite{Khachaturyan-01,Levitas-Basak-2017,Artemev-Wang-Khachaturyan-2000-Acta,Artemev-Jin-Khachaturyan-2001-Acta,Artemev-Slutsker-Roytburd-Acta}) order parameters have been used. Important   requirements for the interpolation functions are formulated in
 \cite{Levitas-preston-PRB-I,Levitas-preston-PRB-II,Levitas-IJP-13,Levitas-Roy-Acta-16,Levitas-IJP-18}. Within the multiphase phase-field approaches, the order parameters should be constrained to some specified surfaces in order to control the transformation paths. To control the transformation paths, various constraint hypersurfaces such as hypersphere \cite{Levitasetal-PRB-twin-13}, planar surfaces \cite{Steinbach-96}, and straight lines \cite{Levitasetal-PRB-15,Levitas-Basak-2017} have been used;  see \cite{Levitas-Basak-2017} for a review. The double-well \cite{Khachaturyan-01,Levitas-Basak-2017,Artemev-Slutsker-Roytburd-Acta,Levitasetal-PRB-15,Levitasetal-PRB-twin-13} or double-obstacle \cite{Stupkiewicz-2017} based thermal energies are usually used. The free energies are considered to be smooth functions of the order parameters, and the transformation strains are accepted as linear \cite{Tuma-Stupkiewicz-16,Tuma-Stupkiewicz-Petryk-18,Stupkiewicz-20,Idesman-Levitas-JMPS-05,Levitas-Id-Pres-PRL-04,Idesman-Levitas-JMPS-05,Levitas-etal-PRL-18,Esfahani-Levitas-AM-20,Babaei-Levitas-JMPS-20} or nonlinear functions \cite{Khachaturyan-01,Levitas-Basak-2017,Levin-Levitas-IJSS-13,Cho-Levitas-12-IJSS,Artemev-Wang-Khachaturyan-2000-Acta,Artemev-Jin-Khachaturyan-2001-Acta} of the order parameters smoothly varying between all the phases, in particular, satisfying some additional requirements \cite{Levitas-preston-PRB-I,Levitas-preston-PRB-II}. First large-strain phase-field theory and computational approaches were presented in  \cite{levitasetal-prl-09,Levitas-IJP-13,Levin-Levitas-IJSS-13}. They utilized the methods of repetitive superposition of large strains, developed by Levin in \cite{Levin-IJSS-98,Levin-Zingerman-ASME-98,Levin-book-99} for viscoelastic materials, extended  to materials with phase transformations.

A  gradient (of the order parameters) based nonlocal energy is considered which introduces a finite interface width between the phases. The time evolution of these order parameters describing the kinetics of the PTs is derived using the laws of  thermodynamics yielding a system of coupled Ginzburg-Landau equations. The interfacial stresses, consisting of the elastic and structural components and which play an important role in the nucleation of the phases and also in their kinetics and growth, have been considered \cite{levitas-javanbakht-PRL-10,Levitas-14a,Levitas-Warren-JMPS-16}. A detailed comparison between these various multiphase phase-field approaches to MTs is presented  in \cite{Levitas-Basak-2017}.

 The multiphase phase-field model for studying the multivariant MTs developed by the authors  in \cite{Levitas-Basak-2017} yields non-contradictory results for a two-variant system. However, for a system with more than two variants, some contradictions have been observed in relation to the gradient energy and the system of kinetic equations. One of the aims of this work is to discuss those issues from that model and present a non-contradictory multiphase phase-field model for MTs. A gradient energy proposed therein simplifies consistently for a two-variant system and matches with the well-established result (see \cite{Levitas-Basak-2017} for the discussion), but a contradiction is observed when the system contains more  than two variants as discussed in Sec. \ref{free_energiess}. An alternative form of  the gradient energy has been used here which has similarities with the gradient energy used in \cite{Steinbach-96,Artemev-Wang-Khachaturyan-2000-Acta,Stupkiewicz-2017,Stupkiewicz-20} and yields non-contradictory results for any number of variants. In the present model, we however multiply this gradient term with the determinant of the total deformation gradient while determining the total system energy to ensure an appropriate form of the structural stress tensor, which was however not considered in \cite{Steinbach-96,Artemev-Wang-Khachaturyan-2000-Acta,Stupkiewicz-2017,Stupkiewicz-20}. Furthermore, we point out in Sec. \ref{kin_laws} that the coupled kinetic equations for the order parameters are non-contradictory for a two-variant system, but leads to contradictions for an $N$-variant system for $N>2$, if the kinetic coefficients are assumed to be constants.  We thus introduce a system of kinetic equations with  kinetic coefficients which are step-wise functions of the order parameters and the driving forces, which is motivated from Ref. \cite{Idesman-Levitas-JMPS-05}. 

\noindent{\em \bf Contribution of the paper}. The contributions of this paper are mainly two-fold:

i) We present a thermodynamically consistent nanoscale phase-field approach for multivariant MTs considering non-contradictory gradient energies and the local energies including the barrier, chemical, and elastic energy, as well as energies penalizing the multiphase junctions, while the deviations of the transformation paths for  $\sf A\leftrightarrow \sf M$ and $\sf M_i\leftrightarrow \sf M_j$ PTs from the specified paths are appropriately controlled. The issues with the existing gradient energy models are discussed. Furthermore, a consistent kinetic model for coupled Ginzburg-Landau equations is derived, and the issues with the existing kinetic models are discussed. The present model can be used for MTs with any number of variants.

ii) The crystallographic solution for the twins within twins microstructure arising in cubic to tetragonal MTs are derived. The evolution and formation of  3D twins within twins microstructures in a single grain is studied using the present phase-field approach. The simulation results are in good agreement with the crystallographic solution and the experimental results.

 \noindent{\em Notations.} The multiplication and the inner product between two arbitrary second order tensors ${\fg A}$ and ${\fg D} $ are denoted by $({\fg A} \cdot{\fg D})_{ab}=A_{ac} D_{cb}$ and  ${\fg A}:{\fg D}=A_{ab} D_{ba}$, respectively, where $A_{ab}$ and $D_{ab}$ are the components of the tensors in a right-handed orthonormal Cartesian basis $\{\fg e_1,\fg e_2,\fg e_3\}$. The repeated indices imply Einstein's summation.  The  Euclidean norm of $\fg A$ is defined by  $|\fg A|=\sqrt{{\fg A}:{\fg A}^T}$.  The second-order identity tensor is denoted by ${\fg I}$.
  $\fg A^T$,  $tr\,\fg A$, $det\,\fg A$,  $sym(\fg A)$, and $skew(\fg A)$ denote the transpose, trace, determinant, symmetric part, and skew part of $\fg A$, respectively. For an invertible tensor $\fg A$, its inverse is denoted by $\fg A^{-1}$. The tensor or dyadic product between two arbitrary vectors $\fg a$ and $\fg b$ is denoted by $\fg a\otimes\fg b$. The reference, stress-free intermediate, and deformed or current configurations are denoted by $\Omega_0$, $\Omega_t$, and $\Omega$, respectively. The volumes in the reference and current configurations are denoted by $V_0$ and $V$, and their external boundaries are denoted by $S_0$ and $S$, respectively. 
 The symbols $\nabla_0(\cdot)$ and $\nabla(\cdot)$ denote the gradient operators in $\Omega_0$ and $\Omega$, respectively. The Laplacian operators in $\Omega_0$ and $\Omega$ are designated by $\nabla_0^2 := \nabla_0 \cdot \nabla_0 $ and $\nabla^2 := \nabla \cdot \nabla$, respectively. The symbol $:=$ implies equality by definition.

 \section{Coupled mechanics and phase-field model}
 \label{system_eqns}
In this section, we describe our  multiphase phase-field model, mostly based on  \cite{Levitas-Basak-2017} but with further development in terms of non-contradictory gradient energy and kinetic relationships between the rate of the order parameters and conjugated thermodynamical driving forces. The free energy used in the model is presented. The coupled elasticity equations and a new system of coupled Ginzburg-Landau equations are derived. A comparison of the present model with the previous models from the literature is also presented.

 \subsection{Order parameters} For the MTs  in a system with austenite and $N$ martensitic variants we consider $N+1$ order
 parameters $\eta_0, \eta_1, \ldots, \eta_i, \eta_j, \ldots,\eta_N$, where  $\eta_0$ describes
 ${\sf A}\leftrightarrow{\sf M}$  transformations such that $\eta_0=0$ in $\sf A$ and $\eta_0=1$ in $\sf M$, and  $\eta_i$ (for $i=1,\ldots,N$) describes the variant ${\sf M}_i$ such that $\eta_i=1$ in ${\sf M}_i$ and $\eta_i=0$ in ${\sf M}_j$ for all $j\neq i$. Such descriptions for the order parameters were introduced by the authors in earlier work \cite{Levitas-Basak-2017}.
 The order parameters  $\eta_1, \,\eta_2,  \ldots, \eta_N$ are constrained to lie on a plane by satisfying (see \cite{Levitas-Basak-2017}
 for details)
 \begin{equation}
 \sum_{i=1}^N \eta_i=1.
  \label{constraintss}
 \end{equation}

\subsection{Kinematics}
 \label{kinematicss1}
 The position vector of a particle in the deformed configuration $\Omega$ at time instance $t$ is given by $\fg r(\fg r_0,t)=\fg r_0+\fg u(\fg r_0,t)$, where $\fg r_0$ is the position vector of that particle  in $\Omega_0$, and $\fg u$ is the displacement vector.  We consider the following multiplicative decomposition for the total deformation gradient $\fg F$ \cite{Levitas-14a}:
  \begin{equation}
{\boldsymbol F}:=\nabla_0\fg r = \fg F_e\cdot\fg F_t=\fg V_e\cdot\fg R\cdot\fg U_t,
 \label{multdecom}
\end{equation}
where the subscripts $e$ and $t$ denote elastic and transformational parts, respectively, and $\fg F_e$ and $\fg F_t$ are  the elastic and transformational parts of $\fg F$. The tensors $\fg F_e$ and $\fg F_t$ are also decomposed into $\fg F_e=\fg V_e\cdot\fg R_e$
and $\fg F_t=\fg R_t\cdot\fg U_t$, respectively, where $\fg V_e$ is the left elastic stretch tensor (symmetric), $\fg U_t$ is the right transformation stretch tensor (symmetric), and $\fg R_e$ and $\fg R_t$ are rotations. In Eq. \eqref{multdecom} we have used $\fg R=\fg R_e\cdot \fg R_t$. We denote $J=det\,\fg F $, $J_t=det\,\fg F_t$, and $J_e=det\,\fg F_e $. Hence, by Eq. (\ref{multdecom}), $J=J_t J_e$.
The Lagrangian total and elastic strain tensors are defined as
\begin{equation}
 {\boldsymbol E} := 0.5(\fg C-{\boldsymbol I}), \quad\text{and}\quad  {\boldsymbol E}_e := 0.5(\fg C_e-{\boldsymbol I}),
   \label{strainsCGL}
 \end{equation}
 respectively, where $\fg C={\boldsymbol F}^T\cdot{\boldsymbol F}$, and $\fg C_e={\boldsymbol F}_e^T\cdot{\boldsymbol F}_e$.
We define another measure of the total and elastic strain tensors  as
 \begin{equation}
 {\boldsymbol b} = 0.5(\fg B-{\boldsymbol I}), \quad\text{and }\quad {\boldsymbol b}_e = 0.5(\fg B_e-{\boldsymbol I}),
   \label{strainsCG}
 \end{equation}
respectively, where $ \fg B = {\boldsymbol F}\cdot{\boldsymbol F}^T=\fg V^2$, $\fg B_e = {\boldsymbol F}_e\cdot{\boldsymbol F}_e^T=\fg V_e^2$, and $\fg V=\sqrt{{\boldsymbol B}}$ is the left total stretch tensor.

\noindent{\em Kinematic model for $\fg F_t$.} We consider $\fg F_t$ as a  linear combination of the Bain strains multiplied  by the  interpolation functions  related to the order parameters \cite{Levitas-Basak-2017}:
 \begin{equation}
{\boldsymbol F}_t={\boldsymbol U}_t =\fg I+ \sum_{i=1}^N (\fg U_{ti}-\fg I) \varphi(a_\varepsilon,\eta_0)\phi_i(\eta_i),
 \label{utilde}
\end{equation}
where $\fg U_{ti}$ is the Bain stretch tensor for $\mathsf M_i$. We take the interpolation functions $\varphi(a_\varepsilon,\eta_0)$ and $\phi_i(\eta_i)$ as \cite{Levitas-Basak-2017}
\begin{eqnarray}
\varphi(a_\varepsilon, \eta_0) &=& a_\varepsilon\eta_0^2+(4-2a_\varepsilon)\eta_0^3 +(a_\varepsilon-3)\eta_0^4,
\quad\text{and} \nonumber\\
 \phi_i(\eta_i) &=& \eta_i^2(3-2\eta_i) \quad\text{for all }i=1,2,\ldots,N,
\label{interpolations}
\end{eqnarray}
respectively, which satisfy the following conditions derived from the requirement of thermodynamic equilibrium of the homogeneous phases:
\begin{eqnarray}
&&\varphi(a_\varepsilon,0) =0, \quad \varphi(a_\varepsilon,1) =1, \quad\text{and}\quad
\frac{\partial\varphi(a_\varepsilon,0)}{\partial\eta_0}= \frac{\partial\varphi(a_\varepsilon,1)}{\partial\eta_0}=0; \nonumber\\
&& \phi_i(0) =0, \quad \phi_i(1) =1, \quad\text{and}\quad
  \frac{\partial\phi_i(\eta_i=0)}{\partial\eta_i}= \frac{\partial\phi_i(\eta_i=1)}{\partial\eta_i}=0 \quad\text{for all }i=1,2,\ldots,N.
\label{phicriterias1}
\end{eqnarray}
The constant $a_\varepsilon$ in Eq. \eqref{interpolations} must be within the range $0\leq a_\varepsilon \leq 6$ \cite{Levitas-14a}.
 \subsection{Free energy}
 \label{free_energiess}
 We assume the Helmholtz free energy per unit mass of the body as  \cite{Levitas-Basak-2017,Levitas-14a}:
 \begin{eqnarray}
 \psi(\fg F,\theta,\eta_0,\eta_i,\nabla\eta_0,\nabla\eta_i) = \psi^l(\fg F_e,\theta,\eta_0,\eta_i)+ J\psi^\nabla(\eta_0,\nabla\eta_0,\nabla\eta_i),
\label{MF0_B1}
\end{eqnarray}
where $ \psi^l(\fg F_e,\theta,\eta_0,\eta_i)$ is the local part of the free energy density and $\psi^\nabla(\eta_0,\nabla\eta_0,\nabla\eta_i)$ is the gradient based nonlocal energy accounting for the  energies of all the interfaces. We have taken $ \psi^l$ as
\begin{eqnarray}
\psi^l(\fg F_e,\theta,\eta_0,\eta_i) &=& \frac{J_t}{\rho_0}\psi_e(\fg F_e,\theta,\eta_0,\eta_i)
+J\breve{\psi}^{\theta}(\theta,\eta_0,\eta_i) +\tilde{\psi}^\theta(\theta,\eta_0,\eta_i) + \psi_p(\eta_0,\eta_i),
\label{MF0_B1s}
\end{eqnarray}
  where $\psi_e$ is the strain energy per unit volume in $\Omega_t$, $\breve\psi^\theta$ is the barrier energy related to
  ${\sf A}\leftrightarrow{\sf M}$ PT and all the variant$\leftrightarrow$variant transformations, $\tilde{\psi}^\theta $ is the
  thermal/chemical energy for ${\sf A}\leftrightarrow{\sf M}$ transformations,  $\psi_p$ penalizes various triple and higher junctions between all the phases and also accounts for the penalization in energies for the deviation of the transformation paths from the assigned ones, $\theta>0$ is the absolute temperature, and $\rho_0$ is the
 density of the solid in $\Omega_0$.  In Eqs. \eqref{MF0_B1} and \eqref{MF0_B1s}, the barrier energy and the gradient energy are multiplied by $J$ following \cite{Levitas-14a} in order to obtain the desired expression for the structural stresses; see Eq. (\ref{surfaceStress2vsg}).
Any material property $B$ for each particle of the body are determined using \cite{Levitas-Basak-2017}
 \begin{equation}
B(\eta_0,\eta_i, \theta, {\boldsymbol F}) = B_0(1-\varphi(a,\eta_0)) + \varphi(a,\eta_0)\sum_{i=1}^N B_i\phi_i (\eta_i),
 \label{properties}
 \end{equation}
where $B_0$ and $B_i$ are the material properties of the homogeneous phases $\mathsf A$ and $\mathsf M_i$, respectively,  $\varphi(a,\eta_0)$
has the same expression as $\varphi(a_\varepsilon,\eta_0)$ given by Eq. (\ref{interpolations})$_1$ when $a_\varepsilon$ is replaced by $a$ therein.

 The expressions for  all the energies introduced in Eqs. \eqref{MF0_B1} and \eqref{MF0_B1s} are given below:

 (i) {\em Strain energy}: We consider $\psi_e$ as \cite{Levitas-Basak-2017}
\begin{equation}
\psi_e =0.5 \fg E_e:\hat{\boldsymbol{\mathcal C}}_e(\eta_0,\eta_i):\fg E_e,
 \label{psi1}
\end{equation}
where the fourth-order elastic modulus tensor at any particle is taken following Eq. \eqref{properties} as \cite{Levitas-Basak-2017}
\begin{equation}
\hat{\boldsymbol{\mathcal C}}_e(\eta_0,\eta_i) = (1-\varphi(a,\eta_0)) \hat{\boldsymbol{\mathcal C}}_{(e)0}
 + \varphi(a,\eta_0)\sum_{i=1}^N \phi_i (\eta_i)  \hat{\boldsymbol{\mathcal C}}_{(e)i},
 \label{psi123}
\end{equation}
and $\hat{\boldsymbol{\mathcal C}}_{(e)0}$ and  $\hat{\boldsymbol{\mathcal C}}_{(e)i}$ are the fourth order elastic modulus tensors of $\sf A$ and
 ${\mathsf M}_i$, respectively.

 (ii) {\em Barrier energy}: The total energy of the barriers between $\sf A$ and $\sf M$ and between all the variants is \cite{Levitas-Basak-2017}
\begin{equation}
\breve{\psi}^{\theta} = A_{0M}\,\eta_0^2 (1-\eta_0)^2
+  \varphi(a_b, \eta_0) \tilde{A}\sum_{i=1}^{N-1}\sum_{j=i+1}^N \eta_i^2\, \eta_j^2 ,
\label{psi2}
\end{equation}
 where $A_{0M}$ and $\tilde{A}$ are the coefficients for the barrier energies between ${\sf A}$ and ${\sf M}$, and ${\sf M}_i$ and ${\sf M}_j$ (for all $i\neq j$), respectively.

 (iii) {\em Thermal energy}: The thermal energy of a particle undergoing ${\sf A}\leftrightarrow{\sf M}$ PTs is taken as \cite{Levitas-preston-PRB-I,Levitas-preston-PRB-II,Levitas-IJP-13,Levitas-Basak-2017}
 \begin{equation}
\tilde{\psi}^\theta =  \psi_{0}^{\theta}(\theta)  + \eta_0^2(3-2\eta_0) \, \Delta \psi^{\theta} (\theta), \quad
\text{where }
\Delta\psi^\theta = -\Delta s_{0M} (\theta-\theta_e),
\label{psi33}
\end{equation}
$\psi_{0}^{\theta}$ is the thermal energy of  $\sf A$, $\Delta \psi^{\theta}=\psi^\theta_M-\psi^\theta_0$ is the thermal energy difference between $\sf A$ and $\sf M$ phases,  $\Delta s_{0M}=s_M-s_0$, $s_0$ and $s_M$ are the entropies of $\sf A$ and $\sf M$, respectively, per unit volume,  and $\theta_e$ is the thermodynamic equilibrium temperature between $\sf A$ and $\sf M$ phases.

 (iv) {\em Gradient energy}: We consider the gradient energy, taking all the interfacial energies into consideration, as
 \begin{equation}
 \psi^\nabla = \frac{\beta_{0M}}{2\rho_0} {\left| {\nabla \eta_0} \right|^2} +\frac{1}{2\rho_0}\tilde\varphi(\eta_0 ,a_\beta, a_c)
\sum_{i=1}^{N-1}\sum_{j=i+1}^N {\beta_{ij}}\nabla\eta_i\cdot\nabla\eta_j,
\label{psi6}
\end{equation}
where  the interpolation function $\tilde\varphi$ is taken as \cite{Levitas-Basak-2017}
\begin{equation}
  \tilde\varphi(a_\beta, a_c,\eta_0) = a_c+{a_\beta }{\eta_0^2} -2[{a_\beta }-2(1-a_c)]{\eta_0^3}+ [{a_\beta }- 3(1-a_c)]{\eta_0^4},
\label{MF0_B1b}
 \end{equation}
where the constant is taken as $0<a_c\ll 1$, and the purpose of considering it in Eq. \eqref{MF0_B1b} is discussed in \cite{Levitas-Basak-2017}.
 In Eq. \eqref{psi6}, $\beta_{0M}$ and $\beta_{ij}$ are the gradient energy coefficients for ${\sf A}$-${\sf M}$ and ${\sf M}_i$-${\sf M}_j$ interfaces, respectively.  In Eq. \eqref{MF0_B1b} $a_\beta$ and $a_c$ are constant parameters. The gradient energy similar to Eq. \eqref{psi6} was earlier used in \cite{Stupkiewicz-20,Steinbach-96}.
Notably, the authors earlier introduced  another form of the gradient energy in \cite{Levitas-Basak-2017}  given by
\begin{eqnarray}
 \psi^\nabla = \frac{1}{2\rho_0} \left[{\beta_{0M}}{\left| {\nabla \eta_0} \right|^2} +
\sum_{i=1}^{N}\sum_{j=1, \neq i}^N \frac{\beta_{ij}}{8} |\nabla \eta_i-\nabla \eta_j|^2\tilde\varphi(\eta_0 ,a_\beta, a_0)\right].
\label{Eq-pot-DGf}
\end{eqnarray}
 This energy given by Eq. \eqref{Eq-pot-DGf} simplifies to
\begin{equation}
 \psi^\nabla =\frac{1}{2\rho_0}{\beta_{0M}}{\left| {\nabla \eta_0} \right|^2} +\frac{\tilde\varphi(\eta_0 ,a_\beta, a_0)}{8\rho_0} \left(\beta_{12}{\left| \nabla \eta_1-\nabla\eta_2 \right|^2}+\beta_{21}{\left| \nabla \eta_2-\nabla\eta_1 \right|^2} \right),
\label{two_var_energ}
\end{equation}
for a system with two variants, where applying the constraint $\eta_1+\eta_2=1$ and $\beta_{12}=\beta_{21}$ due to the symmetry \cite{Levitas-14a} we further simplify it to
\begin{equation}
 \psi^\nabla =\frac{1}{2\rho_0}{\beta_{0M}}{\left| {\nabla \eta_0} \right|^2} +\frac{\tilde\varphi(\eta_0 ,a_\beta, a_0)}{2\rho_0} \beta_{12}{\left| \nabla \eta_1\right|^2},
\label{two_var_energdd}
\end{equation}
which is consistent with the results of earlier models; see e.g.  \cite{Levitas-14a} and the references therein. We would like to mention that the coefficients $\beta_{ij}$ in Eq. \eqref{Eq-pot-DGf} for the variant pairs which are in twin relationships would be much smaller than that for the variant pairs which are not in twin relationships. 
For a system with three variants, Eq. \eqref{Eq-pot-DGf} reduces to
\begin{eqnarray}
 \psi^\nabla &=& \frac{{\beta_{0M}}}{2\rho_0} {\left| {\nabla \eta_0} \right|^2} +\frac{1}{8\rho_0}\left[
 (\beta_{12}+\beta_{13})\left| {\nabla \eta_1} \right|^2+(\beta_{12}+\beta_{23})\left| {\nabla \eta_2} \right|^2
 +(\beta_{13}+\beta_{23})\left| {\nabla \eta_3} \right|^2 -\right.\nonumber\\
 && \left. 2\,(\beta_{12}\nabla \eta_1\cdot\nabla \eta_2+\beta_{23}\nabla \eta_2\cdot\nabla \eta_3+
 \beta_{13}\nabla \eta_1\cdot\nabla \eta_3)\right] \tilde\varphi(\eta_0 ,a_\beta, a_0).
\label{three_var_energ3}
\end{eqnarray}
 Let us consider a region for such a three-variant system where only $\sf M_1$ and $\sf M_2$ coexist and $\sf M_3$ is absent, i.e. $\eta_3=0$ and $\eta_1+\eta_2=1$. The gradient energy given by Eq. \eqref{three_var_energ3} in that region is rewritten by applying these conditions as
\begin{eqnarray}
 \psi^\nabla &=& \frac{{\beta_{0M}}}{2\rho_0} {\left| {\nabla \eta_0} \right|^2} +\frac{1}{8\rho_0}
 (\beta_{23}+\beta_{13}+4\beta_{12})\left| {\nabla \eta_1} \right|^2\tilde\varphi(\eta_0 ,a_\beta, a_0).
\label{three_var_energ3e}
\end{eqnarray}
Obviously, the energy parameters $\beta_{23}$ and $\beta_{13}$ are going to influence  the interfacial energy  $\sf M_1$ and $\sf M_2$ variants,    which is nonphysical; see also \cite{Bollada-2020} for an analysis. We have shown that the gradient energy given by Eq. \eqref{Eq-pot-DGf} yields a nonphysical contribution for an interface between two variants from the gradient coefficients which are not related to that interface, and hence this form of gradient energy is not acceptable. However, the energy given by Eq. \eqref{psi6} is non-contradictory for any number of variants, and hence it is accepted in this paper.

  (v) {\em Penalization energy for junctions}: We penalize the triple and higher junctions between all the phases and the deviations of the transformation paths between the variants using  \cite{Levitas-Basak-2017}
\begin{eqnarray}
 \psi_p &=& \sum_{i=1}^{N-1}\sum_{j=i+1}^N  K_{ij}( \eta_i+\eta_j-1)^2\eta_i^2\eta_j^2
  +[1-\varphi  (a_K, \eta_0)] \sum_{i=1}^{N-1}\sum_{j=i+1}^N K_{0ij} \eta_0^2\eta_i^2\eta_j^2+
  \sum_{i=1}^{N-2}\sum_{j=i+1}^{N-1}\sum_{k=j+1}^N K_{ijk} \eta_i^2\eta_j^2\eta_k^2+\nonumber\\
 &&
 [1-\varphi  (a_K, \eta_0)] \sum_{i=1}^{N-2}\sum_{j=i+1}^{N-1}\sum_{k=j+1}^N K_{0ijk} \eta_0^2\eta_i^2\eta_j^2\eta_k^2+
   \sum_{i=1}^{N-3}\sum_{j=i+1}^{N-2}\sum_{k=j+1}^{N-1}\sum_{l=k+1}^N K_{ijkl} \eta_i^2\eta_j^2\eta_k^2\eta_l^2,
\quad\text{where}
\label{psi4}
\nonumber\\
 K_{ii}&=&K_{0ii}=K_{iji}=K_{iik}= K_{0iji}=K_{0iik}=K_{ijil}=K_{iikl}= K_{ijjl}=K_{ijki}=  K_{ijkk} =0.
 \label{psi5}
\end{eqnarray}
 In Eq. \eqref{psi4}, the parameter $K_{ij}\geq 0$  controls the penalization of the deviation of the ${\sf M}_j\leftrightarrow{\sf M}_i$ transformation
 path  from the straight line $\eta_j+\eta_i=1$ for all $\eta_k=0$ and $k\neq j,i$; the constant coefficients $K_{0ij} \geq 0$, $K_{ijk} \geq 0$, $K_{0ijk}\geq 0$, and $K_{ijkl}\geq 0$ are related to the penalization of the junctions between ${\sf A}$-${\sf M}_i$-${\sf M}_j$, ${\sf M}_i$-${\sf M}_j$-${\sf M}_k$, ${\sf A}$-${\sf M}_i$-${\sf M}_j$-${\sf M}_k$, and ${\sf M}_i$-${\sf M}_j$-${\sf M}_k$-${\sf M}_l$, respectively.

 In the absence of all the penalty terms, i.e. when $K_{ij}=K_{0ij}=K_{0ik}=K_{0jk}=K_{ijk}=K_{0ijk}=0$, we can show that for a martensitic region ($\eta_0=1$) with three variants, say, $\sf M_1$, $\sf M_2$ and $\sf M_3$, the barrier energy (see Eq. \eqref{psi2}) in the middle of the triple junction region, i.e., at the point with $\eta_1=\eta_2=\eta_3=1/3$ is $3\times \bar{A}/81=\bar{A}/27$ which is less than the barrier energy $\bar{A}/16$ at the middle line of any variant-variant interface  (a line with, say, $\eta_1=\eta_2=1/2$ and $\eta_3=0$). Hence when $K_{123}\neq 0$ the total energy at a martensitic particle with $\eta_1=\eta_2=\eta_3=1/3$ is
 \begin{equation}
 \left.E_{TJ}\right|_{\eta_1=\eta_2=\eta_3=1/3}=\frac{\bar{A}}{27}+\frac{K_{123}}{729}.
 \label{juncenger}
  \end{equation} It is to be noted that T\'{o}th et al. \cite{Toth_et_al-2015} and Bollada et al.  \cite{Bollada-2020} considered  barrier energy similar to ours given by Eq. \eqref{psi2} but with a common multiplication factor to incorporate  higher energy at the junction region as compared to the respective interface regions. In this paper, we have, however, followed a different and simpler approach for that purpose, where  we have introduced the penalty terms in the free energy, and by varying the coefficients $K_{0ij},\,K_{0ik},\,K_{0jk},\,K_{ijk},\,K_{0ijk}$ we can control the energy of all the junction regions. For example, by tuning the parameter $K_{123}$ in Eq. \eqref{juncenger}, we can make the barrier energy height at the junction region higher than the barrier energy in the interfacial region. A quantitative comparison between our formulation and the approach in  \cite{Toth_et_al-2015,Bollada-2020} is however not given here.

 \subsection{Governing equations}
 We now present the governing equations. Applying the principle of balance of linear and angular momentum, and the first and second law of thermodynamics, and using an approach similar to  \cite{Levitas-14a,Levitas-Basak-2017}, we derive the mechanical equilibrium equation, and the dissipation inequalities listed below.

 \subsubsection{Mechanical equilibrium equations and stresses}
  The mechanical equilibrium equation is given by  \cite{Levitas-14a,Levitas-Basak-2017}
\begin{equation}
\nabla_0\cdot\fg P =\fg 0 \quad\text{in }\Omega_0, \quad\text{or}\quad\nabla\cdot\fg\sigma=\fg 0 \quad\text{in }\Omega,
\label{MF3}
\end{equation}
where the body forces and inertia are neglected, $\fg P$ is the total first Piola-Kirchhoff stress tensor, and $\fg\sigma$ is the total Cauchy stress tensor which is symmetric. The total stresses are composed of their respective elastic and structural parts \cite{Levitas-Basak-2017}:
\begin{equation}
\fg P = \fg P_e+\fg P_{st}, \quad\text{and}\quad \fg \sigma = \fg \sigma_e+\fg \sigma_{st}.
\label{MF4}
\end{equation}
The elastic stresses are given by \cite{Levitas-Basak-2017,Levitas-14a}
\begin{equation}
 \fg P_e = J_t \fg F_e\cdot\hat{\fg S}_e\cdot\fg F_t^{-T},\quad\text{and}\quad
 \fg\sigma_e=J_e^{-1} \fg F_e\cdot\hat{\fg S}_e\cdot\fg F_e^T,
  \label{surfaceStress2vel}
\end{equation}
where $ \hat{\fg S}_e=\frac{\partial\psi_e(\fg E_e)}{\partial\fg E_e}$. For an  isotropic elastic response, $\fg P_e$ and $\fg \sigma_e$ can alternatively be expressed as
\begin{equation}
\fg P_e = J_t\fg V_e^2\cdot\frac{\partial \psi_e(\fg b_e)}{\partial\fg b_e }\cdot\fg F^{-T}, \quad \text{and}\quad
 \fg\sigma_e = J^{-1}_e \fg V_e^2\cdot\frac{\partial \psi_e(\fg b_e)}{\partial\fg b_e },
\label{hl5f34s}
\end{equation}
respectively.
 The general forms of  structural stress tensors are given by  \cite{Levitas-Basak-2017,Levitas-14a}
\begin{eqnarray}
  {\boldsymbol P}_{st} &=&  J\rho_0(\breve{\psi}^{\theta}+\psi^\nabla){\fg F}^{-T}
 -J\rho_0  \left(\nabla\eta_0 \otimes\frac{\partial\psi^\nabla}{\partial\nabla\eta_0}+
  \sum_{i=1}^N  \nabla\eta_i \otimes\frac{\partial\psi^\nabla}{\partial\nabla\eta_i}\right)\cdot\fg F^{-T}, \quad\text{and}
  \nonumber\\
  \fg\sigma_{st} &=&\rho_0(\breve{\psi}^{\theta}+\psi^\nabla)\fg I-\rho_0\left(\nabla\eta_0\otimes
 \frac{\partial\psi^\nabla}{\partial\nabla\eta_0}+ \sum_{i=1}^N \nabla\eta_i\otimes
 \frac{\partial\psi^\nabla}{\partial\nabla\eta_i}\right).
 \label{hl4fjgffg}
\end{eqnarray}
Using the gradient energy given by Eq. \eqref{psi6} in Eq. \eqref{hl4fjgffg}$_{1,2}$, the exact form of the structural stresses are obtained as
\begin{eqnarray}
{\boldsymbol P}_{st} &=& J\rho_0(\breve{\psi}^{\theta}+\psi^\nabla){\fg F}^{-T}-J\beta_{0M}\nabla\eta_0\otimes\nabla\eta_0
\cdot{\fg F}^{-T}- J\tilde\varphi\left(\sum_{i=1}^{N-1}\sum_{j=i+1}^N\beta_{ij}\nabla\eta_i\otimes\nabla\eta_j\right)
\cdot {\fg F}^{-T}, \quad\text{and} \nonumber\\
\fg\sigma_{st} &=& \rho_0(\breve{\psi}^{\theta}+\psi^\nabla)\fg I-\beta_{0M}\nabla\eta_0\otimes\nabla\eta_0
-\tilde\varphi\left(\sum_{i=1}^{N-1}\sum_{j=i+1}^N\beta_{ij} \nabla\eta_i\otimes\nabla\eta_j\right).
 \label{surfaceStress2vsg}
\end{eqnarray}

\subsection{Dissipation inequality and Ginzburg-Landau equations}
\label{kin_laws}
The dissipation inequalities for the order parameter $\eta_0$ and also for the order parameter $\eta_1,\ldots,\eta_N$  are obtained as (see  \cite{Levitas-Basak-2017}   for the details)
\begin{equation}
D_0= \dot\eta_0 X_0\geq 0 \quad\text{and}\quad   D_M=\sum_{i=1}^N\dot\eta_iX_i\geq 0,
 \label{dissiinea}
\end{equation}
respectively, where the conjugate `forces' $X_0$ and $X_i$ correspond to the `fluxes' $\dot\eta_0$ and $\dot\eta_i$, respectively, are given by
\begin{eqnarray}
X_k = -\rho_0\frac{\partial\psi}{\partial\eta_k}+\nabla_0\cdot\left( \rho_0 J\frac{\partial\psi^\nabla}{\partial\nabla_0\eta_k} \right) \quad \text{for all } k=0,1,\ldots,N.
\label{forcessbox1s}
\end{eqnarray}
Using Eqs. \eqref{MF0_B1} and \eqref{MF0_B1s} in \eqref{forcessbox1s} and also applying the following identities (for all $k=0,1,\ldots,N$)
\begin{equation}
\nabla_0\eta_k=\fg F^t\cdot\nabla\eta_k,
\label{identity1}
\end{equation}
\begin{equation}
\displaystyle\frac{\partial\psi^\nabla}{\partial\nabla_0\eta_k}=\fg F^{-1}\cdot\frac{\partial\psi^\nabla}{\partial\nabla\eta_k},
\label{identity2}
\end{equation}
which can be easily proved using the indicial notations, we get the conjugate forces $X_0$ and $X_i$ (for all $i=1,\ldots,N$) as
\begin{eqnarray}
X_0 &=& \left(\fg P_e^T\cdot\fg F-J_t\psi_e\fg I\right):\fg F_t^{-1}\cdot\frac{\partial\fg F_t}{\partial\eta_0}-
J_t \left.\frac{\partial\psi_e}{\partial\eta_0}\right|_{\fg F_e}-
\rho_0(6\eta_0-6\eta_0^2)\Delta\psi^\theta-J\rho_0\tilde{A}\sum_{i=1}^{N-1}
\sum_{j=i+1}^N\eta_i^2\eta_j^2\frac{\partial\varphi(a_b,\eta_0)}{\partial\eta_0}-
\nonumber\\
&&  J\rho_0 [A_{0M}(\theta)+(a_\theta-3)\Delta\psi^\theta(\theta)] (2\eta_0-6\eta_0^2+4\eta_0^3)
-\frac{J}{2}\frac{\partial\tilde\varphi(a_\beta,a_c,\eta_0)}
{\partial\eta_0}\sum_{i=1}^{N-1}\sum_{j=i+1}^N\beta_{ij}(\fg C^{-1}\cdot\nabla_0\eta_i)\cdot\nabla_0\eta_j-
\nonumber\\
&&
\rho_0\left(\sum_{i=1}^{N-1}\sum_{j=i+1}^N K_{0ij}\eta_i^2\eta_j^2+ \sum_{i=1}^{N-2}\sum_{j=i+1}^{N-1}
\sum_{k=j+1}^N K_{0ijk}\eta_i^2\eta_j^2\eta_k^2\right)
\left[2(1-\varphi(a_K,\eta_0))\eta_0-\frac{\partial\varphi(a_K,\eta_0)}{\partial\eta_0}\eta_0^2\right]+ \nonumber\\
&& \nabla_0\cdot\left(J\beta_{0M}\nabla_0\eta_0\right);
\label{forcessbox1}
\end{eqnarray}
\begin{eqnarray}
X_i &=& \left(\fg P_e^T\cdot\fg F-J_t\psi_e\fg I\right):\fg F_t^{-1}\cdot\frac{\partial\fg F_t}{\partial\eta_i}
-J_t \left.\frac{\partial\psi_e}{\partial\eta_i}\right|_{\fg F_e}
-2J\rho_0\tilde{A}\sum_{j=1,\neq i}^N\eta_i\eta_j^2\varphi(a_b,\eta_0)-2\rho_0 \sum_{j=1}^N K_{ij}(\eta_i+\eta_j-1)\times
\nonumber\\
 && (2\eta_i+\eta_j-1)\eta_j^2\eta_i -2\rho_0\left(\sum_{j=1}^N K_{0ij}\eta_j^2
 +\sum_{j=1}^{N-1}\sum_{k=j+1}^N K_{0ijk}\eta_j^2\eta_k^2 \right) \eta_0^2\eta_i(1-\varphi(a_K,\eta_0))-
 \nonumber\\
 &&
2\rho_0 \sum_{j=1}^{N-1}\sum_{k=j+1}^N K_{ijk}\eta_i\eta_j^2\eta_k^2
- 2\rho_0\sum_{j=1}^{N-2}\sum_{k=j+1}^{N-1}\sum_{l=k+1}^N K_{ijkl}\eta_i\eta_j^2\eta_k^2\eta_l^2 +
 \nonumber\\
 && \nabla_0\cdot\left(\tilde\varphi(a_\beta,a_c,\eta_0)J\sum_{j=1}^N
  \beta_{ij}\nabla_0\eta_j\right) \quad
  \text{for all }i=1,2,3,\ldots,N.
\label{forcessxibox1}
\end{eqnarray}
In Eqs. \eqref{forcessbox1} and \eqref{forcessxibox1} the conjugate forces are expressed with respect to the field variables in the reference configuration $\Omega_0$.

Alternatively, these forces can be expressed with respect to the field variables in $\Omega$ as follows. Using the following identity
  (see e.g. Chapter 2 of \cite{Jog-2007})
\begin{equation}
\displaystyle\nabla_0\cdot(Cof\, \fg F)=\nabla_0\cdot(J\fg F^{-t})=\fg 0,
\label{identity3}
\end{equation}
we can rewrite the conjugate force $X_l$ given by Eq. \eqref{forcessbox1s} in terms of the derivative with respect to $\Omega(t)$ as
\begin{eqnarray}
X_k =-J\rho\frac{\partial\psi}{\partial\eta_k}+J\nabla\cdot\left( \rho_0 \frac{\partial\psi^\nabla}{\partial\nabla\eta_k} \right) =J\left[-\frac{\partial(\rho\psi)}{\partial\eta_k}+\nabla\cdot\left( \frac{\partial( \rho J \psi^\nabla)}{\partial\nabla\eta_k} \right)\right] \quad \text{for all } k=0,1,\ldots,N .
\label{forcessbox1g}
\end{eqnarray}
The conjugate forces given by Eqs. \eqref{forcessbox1} and \eqref{forcessxibox1} can be rewritten in terms of the Cauchy stress and the gradient operator   $\nabla(\cdot)$ as
\begin{eqnarray}
X_0 &=& \left(J \fg F^{-1}\cdot\fg \sigma_e\cdot\fg F-J_t\psi_e\fg I\right):\fg F_t^{-1}\cdot\frac{\partial\fg F_t}{\partial\eta_0}-
J_t \left.\frac{\partial\psi_e}{\partial\eta_0}\right|_{\fg F_e}-
\rho_0(6\eta_0-6\eta_0^2)\Delta\psi^\theta-J\rho_0\tilde{A}\sum_{i=1}^{N-1}
\sum_{j=i+1}^N\eta_i^2\eta_j^2\frac{\partial\varphi(a_b,\eta_0)}{\partial\eta_0}-
\nonumber\\
&&  J\rho_0 [A_{0M}(\theta)+(a_\theta-3)\Delta\psi^\theta(\theta)] (2\eta_0-6\eta_0^2+4\eta_0^3)
-\frac{J}{2}\frac{\partial\tilde\varphi(a_\beta,a_c,\eta_0)}
{\partial\eta_0}\sum_{i=1}^{N-1}\sum_{j=i+1}^N\beta_{ij}\nabla\eta_i\cdot\nabla\eta_j-
\nonumber\\
&&
\rho_0\left(\sum_{i=1}^{N-1}\sum_{j=i+1}^N K_{0ij}\eta_i^2\eta_j^2+ \sum_{i=1}^{N-2}\sum_{j=i+1}^{N-1}
\sum_{k=j+1}^N K_{0ijk}\eta_i^2\eta_j^2\eta_k^2\right)
\left[2(1-\varphi(a_K,\eta_0))\eta_0-\frac{\partial\varphi(a_K,\eta_0)}{\partial\eta_0}\eta_0^2\right]+ \nonumber\\
&& J\nabla\cdot\left(\beta_{0M}\nabla\eta_0\right);
\label{forcessbox1gm}
\end{eqnarray}
\begin{eqnarray}
X_i &=& \left(J \fg F^{-1}\cdot\fg \sigma_e\cdot\fg F-J_t\psi_e\fg I\right):\fg F_t^{-1}\cdot\frac{\partial\fg F_t}{\partial\eta_i}
-J_t \left.\frac{\partial\psi_e}{\partial\eta_i}\right|_{\fg F_e}
-2J\rho_0\tilde{A}\sum_{j=1,\neq i}^N\eta_i\eta_j^2\varphi(a_b,\eta_0)-2\rho_0 \sum_{j=1}^N K_{ij}(\eta_i+\eta_j-1)\times
\nonumber\\
 && (2\eta_i+\eta_j-1)\eta_j^2\eta_i -2\rho_0\left(\sum_{j=1}^N K_{0ij}\eta_j^2
 +\sum_{j=1}^{N-1}\sum_{k=j+1}^N K_{0ijk}\eta_j^2\eta_k^2 \right) \eta_0^2\eta_i(1-\varphi(a_K,\eta_0))-
 \nonumber\\
 &&
2\rho_0 \sum_{j=1}^{N-1}\sum_{k=j+1}^N K_{ijk}\eta_i\eta_j^2\eta_k^2
- 2\rho_0\sum_{j=1}^{N-2}\sum_{k=j+1}^{N-1}\sum_{l=k+1}^N K_{ijkl}\eta_i\eta_j^2\eta_k^2\eta_l^2 +
 \nonumber\\
 && J \nabla\cdot\left(\tilde\varphi(a_\beta,a_c,\eta_0)\sum_{j=1}^N
  \beta_{ij}\nabla\eta_j\right) \quad
  \text{for all }i=1,2,3,\ldots,N.
\label{forcessxibox1g}
\end{eqnarray}

 From the dissipation inequality \eqref{dissiinea}$_1$ we derive the kinetic law for $\eta_0$  as
\begin{equation}
\dot\eta_0 = L_{0M} X_0,
 \label{dissiineq}
\end{equation}
where $ L_{0M}>0$ is the kinetic coefficient for $\sf A\leftrightarrow\sf M$ PTs. In order to derive the kinetic laws for the order parameters $\eta_1,\ldots,\eta_N$ using the inequality \eqref{dissiinea}$_2$, we introduce
\begin{equation}
\dot\eta_i = \sum_{j=1}^N\dot\eta_{ij}, \qquad\text{where }\dot\eta_{ij}=-\dot\eta_{ji} \text{ and } \dot\eta_{ii}=0 \quad \text{for all } i,j=1,\ldots,N.
 \label{vari_toj}
\end{equation}
 Using Eq. \eqref{vari_toj}, the dissipation rate due to the evolution of the martensitic variants given by Eq. \eqref{dissiinea}$_2$ is rewritten as
\begin{eqnarray}
  D_M &=& \sum_{i=1}^N\dot\eta_iX_i = \sum_{i=1}^N\sum_{j=1}^NX_i\dot\eta_{ij} \qquad \text{(using Eq. \eqref{vari_toj}$_1$)}\nonumber\\
  &=& \sum_{i=1}^N\sum_{j=1}^N X_{ij}\dot\eta_{ij}
  +\sum_{i=1}^N\sum_{j=1}^N X_j\dot\eta_{ij} \qquad \text{(using $X_{ij}=X_i-X_j$)}\nonumber\\
  &=&  \sum_{i=1}^N\sum_{j=1}^N X_{ij}\dot\eta_{ij}- \sum_{i=1}^N\sum_{j=1}^N X_j\dot\eta_{ji} \qquad \text{(using Eq. \eqref{vari_toj}$_2$)}\nonumber\\
  &=&\sum_{i=1}^N\sum_{j=1}^N X_{ij}\dot\eta_{ij}- \sum_{i=1}^N\sum_{j=1}^N X_i\dot\eta_{ij} \qquad \text{(swapping the indices in the second term)}\nonumber\\
  &=&\sum_{i=1}^N\sum_{j=1}^N X_{ij}\dot\eta_{ij}- \sum_{i=1}^N X_i\dot\eta_i.
 \label{dissiinedd}
\end{eqnarray}
Noticing that the second term on the right-hand side of Eq. \eqref{dissiinedd} is equal to $D_M$ (compare with Eq. \eqref{dissiinea}$_2$) we obtain
\begin{eqnarray}
  D_M &=& \frac{1}{2}\sum_{i=1}^N\sum_{j=1}^N X_{ij}\dot\eta_{ij} = \sum_{j=1}^{N-1}\sum_{i=j+1}^N X_{ij}\dot\eta_{ij} \geq 0.
 \label{dissiinedd1}
\end{eqnarray}
Using the inequality \eqref{dissiinedd1} we derive the Ginzburg-Landau equations for the evolution of the variants as
\begin{eqnarray}
\dot\eta_{ij} = L_{ij}(X_i-X_j) ,
\label{kiness}
\end{eqnarray}
where $L_{ij} \geq 0$ is the kinetic coefficient for transformations between $\sf M_i$ and $\sf M_j$ and is taken as in \cite{Idesman-Levitas-JMPS-05}
\begin{eqnarray}
    L_{ij}  \left\{\begin{array}{@{}lr@{}}
        \neq 0 & \text{if }(X_i-X_j)\geq 0 \quad \text{and} \quad \{0\leq \eta_i<1 \,\,\&\,\, 0<\eta_j\leq 1\}\\
                 \neq 0               & \text{if }(X_i-X_j)\leq 0 \quad \text{and} \quad \{0< \eta_i\leq 1 \,\,\&\,\, 0\leq \eta_j <1\} \\
        =0 & \text{if }(X_i-X_j)\geq 0 \quad \text{and} \quad \{\eta_i=1 \,\,\text{or}\,\, \eta_j=0\}\\
       =0 & \text{if }(X_i-X_j)\leq 0 \quad \text{and} \quad \{ \eta_i=0 \,\,\text{or}\,\,  \eta_j=1\}.
        \end{array}\right.
\label{kinesss}
\end{eqnarray}
 Substituting Eq. \eqref{kiness} in Eq. \eqref{vari_toj} the Ginzburg-Landau equations for all $N$ order parameters $\eta_1,\ldots,\eta_N$ are obtained as
\begin{eqnarray}
 \dot\eta_i = \sum_{j=1, j\neq i}^NL_{ij}(X_i-X_j).
\label{kinessg}
\end{eqnarray}
We note that $L_{ij}$ in Eq. \eqref{kinessg} and defined in Eq. \eqref{kinesss} are piece-wise  constant, jumping between their finite values and zero  depending  on the  driving forces and the order parameters. If $L_{ij}$ is simply assumed to be constants similar to our earlier work in \cite{Levitas-Basak-2017}, an issue would arise. To understand it clearly, let us consider, without any loss of generality, a three-variant martensitic system (where $\eta_0=1$)  with $\sf M_1$, $\sf M_2$, and $\sf M_3$. Using the constraint $\eta_1+\eta_2+\eta_3=1$, the two independent Ginzburg-Landau equations from Eq. \eqref{kinessg}, when expressed for  $\dot\eta_1$ and $\dot\eta_2$, are given by
 \begin{eqnarray}
\dot\eta_1 =L_{12}(X_1-X_2)+L_{13}(X_1-X_3) \quad \text{and} \quad \dot\eta_2 =L_{12}(X_2-X_1)+L_{23}(X_2-X_3).
\label{kinessgp}
\end{eqnarray}
We now consider a martensitic region in the domain where $\sf M_3$ is absent and only the variants $\sf M_1$ and $\sf M_2$ are evolving within an arbitrary time interval. The order parameters $\eta_1$ and $\eta_2$ hence must be determined using the equation $\dot\eta_1=\dot\eta_2 = L_{12}(X_1-X_2)$ as $\eta_1+\eta_2=1$ therein, and this is possible if and only if $L_{13}=L_{23}=0$ therein within that time interval. However, if the coefficients $L_{13}$ and $L_{23}$ are taken as nonzero constants, the contributions from the terms $L_{13}(X_1-X_3)$ and $L_{23}(X_2-X_3)$ would be there unwantedly, since the driving forces  $X_1-X_3$ and $X_2-X_3$ might be nonzero there. The desired condition can obviously be fulfilled by  $L_{ij}$ given by Eq. \eqref{kinesss}, but not by constant $L_{ij}$ considered in \cite{Levitas-Basak-2017}.
The essence of the third and fourth conditions in Eq. \eqref{kinesss} is that if variant $i$ is absent, it cannot be transformed into other variants
\cite{Idesman-Levitas-JMPS-05}.

\subsection{Boundary conditions}
The boundary conditions for the phase-field equations and the mechanics problem namely the Dirichlet, Neumann, and periodic BCs are listed here.

 \noindent{\em Phase-field problem.}
 We have applied the periodic BC for all the order parameters.
 If two boundaries  $S_{p1\eta_k}\subset S_0$ and $S_{p2\eta_k}\subset S_0$ (where $S_{p1\eta_k}\cap S_{p2\eta_k}$ is empty), having opposite unit normals in $\Omega_0$, i.e. $({\fg n_0})_{S_{p1\eta_k}}=-(\fg n_0)_{S_{p2\eta_k}}$,  are subjected to a periodic BC related to the order parameters $\eta_k$ (for $k=0,1,2,\ldots,N$), then the order parameters and their gradients must satisfy
 \begin{equation}
\eta_k |_{S_{p1\eta_k}} =\eta_k |_{S_{p2\eta_k}} \quad \text{and}\quad (\nabla_0\eta_k\cdot\fg n_0 )_{S_{p1\eta_k}} = (\nabla_0\eta_k\cdot\fg n_0 )_{S_{p2\eta_k}}\qquad\text{for all }k=0,1,2,\ldots,N .
  \label{PBC1}
 \end{equation}

  \noindent{\em Mechanics problem.} On the traction boundary $S_{0T}\subset S_0$, the traction vector  is specified (denoted by $\fg p^{sp}$), and on the displacement boundary  $S_{0u}\subset S_0$, the displacements are specified (denoted by $\fg u^{sp}$), i.e.
 \begin{equation}
  \fg P \cdot \fg n_0 = \fg p^{sp} \qquad\text{on }S_{0T},
  \label{traction_sp}
 \end{equation}
 \begin{equation}
 \fg u = \fg u^{sp} \qquad\text{on }S_{0u}.
  \label{traction_spu}
 \end{equation}
 If two boundaries  $S_{pu1}\subset S_0$ and $S_{pu2}\subset S_0$ (where $S_{pu1}\cap S_{pu2}$ is empty) are subjected to a periodic BC on the displacements $\fg u$, then the displacements on these boundaries are related by
 \begin{equation}
{\fg u}|_{S_{pu1}} ={\fg u}|_{S_{pu2}}+({\fg F}_h-\fg I)\cdot{\fg r}_0,
  \label{PBC2}
 \end{equation}
 where ${\fg F}_h$ is a specified homogeneous deformation gradient.
 The  mixed boundary conditions where, on a single surface, some components of displacements are specified and some
 components of the traction are specified, are also used. Furthermore, one or more displacement components on a boundary may be related by the periodic BC  given by Eq. \eqref{PBC2} with another appropriate boundary.

\section{Crystallographic solutions for  twins within twins microstrucrure  }
\label{crystTh}
\begin{figure}[t!]
\centering
  \includegraphics[width=3.5in, height=3.2in] {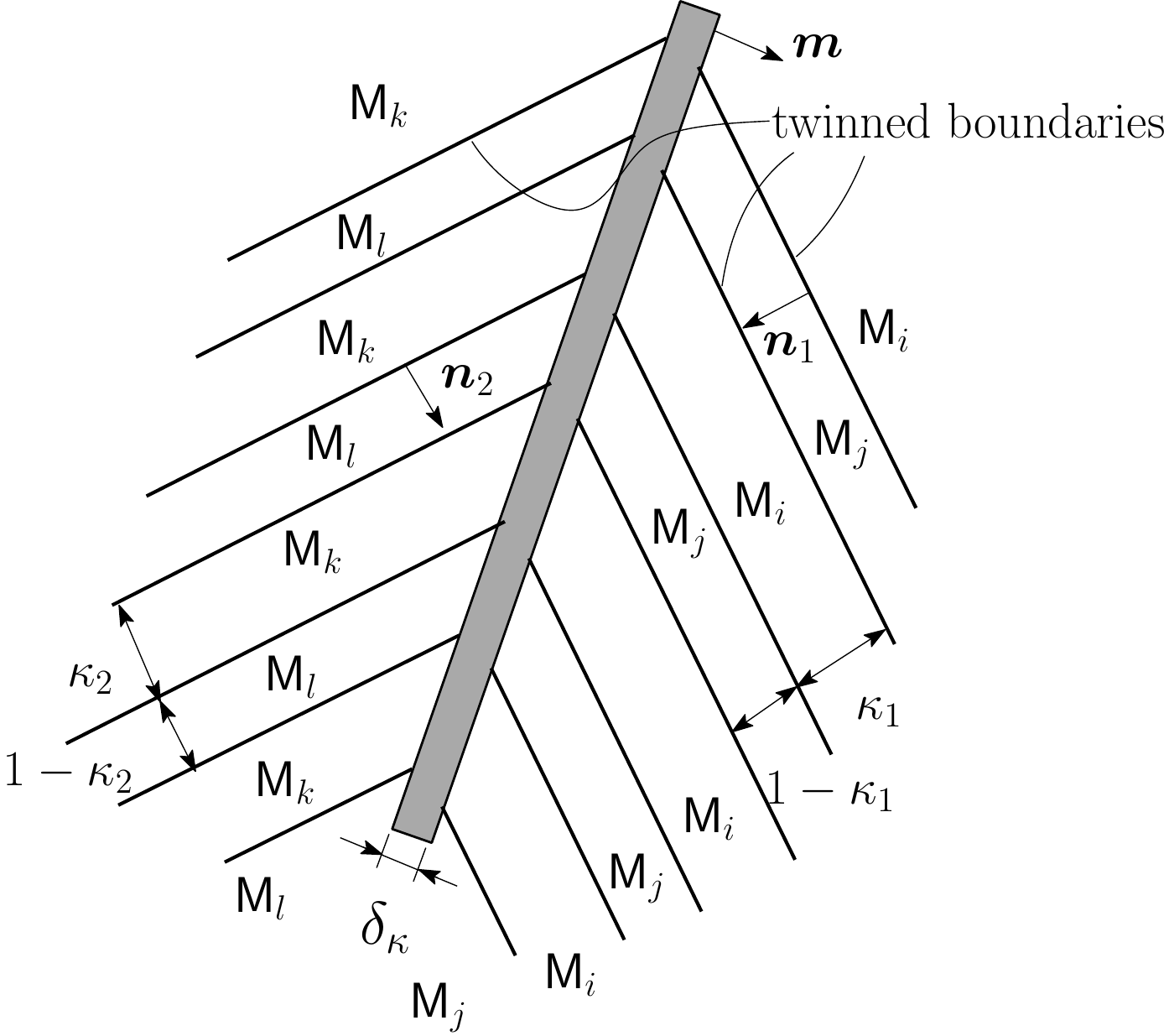}
\caption{A schematic of twins within twins.}
\label{twin_twin}
\end{figure}
In this section, we obtain the analytical solutions for the twins within twins microstructures for the cubic to tetragonal MTs using  the crystallographic theory (see e.g. \cite{Bha04}).  A schematic of twins within twins is shown in Fig. \ref{twin_twin} where the twins formed by a variants pair $\sf M_i$ and $\sf M_j$, and the twins formed by another variants pair  $\sf M_k$ and $\sf M_l$ form an interface of finite thickness $\delta_k$ shown by a shaded region. The volume fractions of $\sf M_i$ and $\sf M_k$ in the respective twins are $\kappa_1$ and $\kappa_2$. The variants $\sf M_i$ and $\sf M_j$ need not be in twin relationship with the other two variants $\sf M_k$ or $\sf M_l$ \cite{Bha04}.

\subsection{Crystallographic equations and general solutions}
\noindent{\bf Crystallographic equations for twin-twin:} The equations for the twins between
the pairs $\sf M_i$-$\sf M_j$, and $\sf M_k$-$\sf M_l$ are (see e.g. Chapter 7 of \cite{Bha04} and \cite{Chu_thesis1993,Abeyaratne-Chu_96})
\begin{equation}
\fg R_2\cdot\fg U_{tj}-\fg R_1\cdot\fg U_{ti} = {\fg a}_1'\otimes \fg n_1, \quad \text{and} \quad
\fg R_4\cdot\fg U_{tl}-\fg R_3\cdot\fg U_{tk} = {\fg a}_2'\otimes \fg n_2,
  \label{gleqsbcgg1}
 \end{equation}
 respectively, where $\fg R_1$, $\fg R_2$, $\fg R_3$,  and $\fg R_4$ are the rotation tensors; $\fg n_1$ and $\fg n_2$ are the unit normals to the respective twin boundaries such that $\fg n_1$ points into $\sf M_j$ and $\fg n_2$ points into $\sf M_l$ (see Fig. \ref{twin_twin}); the vectors ${\fg a}_1'$ and ${\fg a}_2'$ are related to the simple shear deformations. The governing equation for the twins within twins shown in Fig. \ref{twin_twin} is (Chapter 7 of \cite{Bha04} and \cite{Chu_thesis1993,Abeyaratne-Chu_96})
 \begin{equation}
(\kappa_1\fg R_2\cdot\fg U_{tj} +(1-\kappa_1)\fg R_1\cdot\fg U_{ti})-(\kappa_2\fg R_4\cdot\fg U_{tl} +(1-\kappa_2)\fg R_3\cdot\fg U_{tk})= {\fg b}'\otimes  {\fg m},
  \label{gleqsbcgg3}
 \end{equation}
 where $\kappa_1$ and $\kappa_2$ are the volume fractions of $\sf M_i$ and $\sf M_k$ in the respective twins, ${\fg m}$ is the unit normal to the twin-twin interface shown in the figure, and ${\fg b}'$ is a vector related to the deformation.

 Equations \eqref{gleqsbcgg1}$_1$ and \eqref{gleqsbcgg1}$_2$ can be rewritten as
 \begin{equation}
 \fg Q_1\cdot\fg U_{tj}-\fg U_{ti} = \fg a_1\otimes \fg n_1, \quad \text{and} \quad
   \qquad \fg Q_2\cdot\fg U_{tl}-\fg U_{tk} = \fg a_2\otimes \fg n_2,
  \label{gleqsbcgg2}
 \end{equation}
 where $\fg Q_1=\fg R_1^t\cdot\fg R_2, $ $\fg a_1= \fg R_1^t\cdot\fg a_1'$, $\fg Q_2=\fg R_3^t\cdot\fg R_4$,  and $ \fg a_2= \fg R_3^t\cdot\fg a_2'$. Similarly, using Eq. \eqref{gleqsbcgg2}$_{1,2}$, Eq. \eqref{gleqsbcgg3} is rewritten as
  \begin{equation}
\fg Q_3\cdot(\fg U_{ti} +\kappa_1 {\fg a}_1\otimes \fg n_1)-(\fg U_{tk} +\kappa_2 {\fg a}_2\otimes \fg n_2)= {\fg b}\otimes \fg m,
  \label{gleqsbcgg31}
 \end{equation}
 where $\fg Q_3 = \fg R_3^t\cdot\fg R_1$ and $\fg b = \fg R_3^t\cdot{\fg b}'$. In order to solve Eq. \eqref{gleqsbcgg31}, we
post-multiply it with $(\fg U_{tk} +\kappa_2 {\fg a}_2\otimes \fg n_2)^{-1}$ and rearrange the terms to rewrite the equation as \cite{Chu_thesis1993}
 \begin{equation}
\fg Q_3\cdot\tilde{\fg A}=\fg I+ {\fg b}\otimes\tilde{ \fg m},
  \label{gleqsbcgg32}
 \end{equation}
where
 \begin{equation}
 \tilde{\fg A}=(\fg U_{ti} +\kappa_1 {\fg a}_1\otimes \fg n_1)\cdot(\fg U_{tk} +\kappa_2 {\fg a}_2\otimes \fg n_2)^{-1}, \quad \text{ and }\quad
 \tilde{\fg m}=(\fg U_{tk} +\kappa_2 {\fg a}_2\otimes \fg n_2)^{-t}\cdot{\fg m}.
  \label{gleqsbcgg34b}
 \end{equation}
The variables to be determined from the above equations \eqref{gleqsbcgg2} to \eqref{gleqsbcgg34b} are $\kappa_1$, $\kappa_2$, $\delta_\kappa$, $\fg a_1$, $\fg a_2$, $\fg n_1$, $\fg n_2$, $\fg b$, $\fg m$, $\fg Q_1$, $\fg Q_2$, and $\fg Q_3$.

 \noindent{\bf Twins within twins solution:}
 The solution for Eq. \eqref{gleqsbcgg2}$_1$ and \eqref{gleqsbcgg2}$_2$ are well known (see e.g. Chapter 5 of \cite{Bha04}) and is enlisted here for  completeness. To do so, we define a symmetric tensor $\fg G_1=\fg U_{tj}^{-1}\cdot\fg U_{ti}^2\cdot\fg U_{tj}^{-1}$   corresponding to Eq. \eqref{gleqsbcgg2}$_1$. The  eigenvalues of $\fg G_1$ are denoted by $\lambda_1$, $\lambda_2$, and $\lambda_3$, which are all positive, and the corresponding normalized eigenvectors are denoted by $\fg i_1$, $\fg i_2$, and $\fg i_3$, respectively.  Equation \eqref{gleqsbcgg2}$_1$  has a solution if and only if  $\lambda_1\leq 1$, $\lambda_2= 1$, and $\lambda_3\geq 1$ (assuming $\lambda_1\leq \lambda_2\leq \lambda_3$). The expressions for $\fg a_1$ and $\fg n_1$ are given by
 \begin{eqnarray}
\fg a_1 &=&\zeta_1 \left(\sqrt{\frac{\lambda_3(1-\lambda_1)}{\lambda_3-\lambda_1} } \,\,\fg i_1+\xi\sqrt{\frac{\lambda_1(\lambda_3-1)}{\lambda_3-\lambda_1}}\,\,  \fg i_3\right),\nonumber\\
\fg n_1 &=&\frac{\sqrt{\lambda_3}-\sqrt{\lambda_1}}{\zeta_1\sqrt{\lambda_3-\lambda_1}} \left(-\sqrt{{1-\lambda_1} } \,\,\fg U_{tj}\fg i_1+\xi\sqrt{\lambda_3-1}\,\,  \fg U_{tj}\fg i_3\right)\,
  \label{gleqsbcgg4}
 \end{eqnarray}
where $\xi=\pm 1$, and $\zeta_1$ is such that $|\fg n_1|=1$. The solutions $\fg a_2$ and $\fg n_2$ for the twins between $\sf M_k$ and $\sf M_l$ are similarly obtained using the eigenpairs of $\fg U_{tl}^{-1}\cdot\fg U_{tk}^2\cdot\fg U_{tl}^{-1}$ in Eq. \eqref{gleqsbcgg4}. The rotations $\fg Q_1$ and $\fg Q_2$ can then be obtained using Eqs. \eqref{gleqsbcgg2}$_{1,2}$.

We now solve the twins within twins equation \eqref{gleqsbcgg32} which has got a form similar to the $\sf A$-twinned martensite interface equation  using the procedure of Ball and James \cite{Ball-James-87} (also see Chapter 7 of \cite{Chu_thesis1993}). Noticing that the Bain stretches $\fg U_{ti}$, $\fg U_{tj}$, $\fg U_{tk}$, and $\fg U_{tl}$ are given for a material, we first obtain $\fg a_1$, $\fg a_2$, $\fg n_1$, and $\fg n_2$ using, e.g., Eq. \eqref{gleqsbcgg4}.  The procedure for obtaining the remaining unknowns $\kappa_1$, $\kappa_2$, $\fg Q_3$, $\fg b$, and $\tilde{\fg m}$ is derived here.
In obtaining these unknowns, let us first assume that the parameters $\kappa_1$ and $\kappa_2$ are given, and solve for the remaining unknown variables.  We introduce
\begin{equation}
\fg G_2 = \tilde{\fg A}^t\cdot\tilde{\fg A},
\label{c2tensor}
\end{equation}
 which is obviously symmetric and positive-definite, and all the eigenvalues are  positive numbers which we denote by $\Lambda_1$, $\Lambda_2$, and $\Lambda_3$. The corresponding normalized eigenvectors are denoted by $\fg j_1$, $\fg j_2$, and $\fg j_3$, respectively.  Equation \eqref{gleqsbcgg32} has a solution if and only if  $\Lambda_1\leq 1$, $\Lambda_2= 1$, and $\Lambda_3\geq 1$ assuming $\Lambda_1\leq \Lambda_2\leq \Lambda_3$. The solutions for the vectors $\fg b$ and $\tilde{\fg m}$ are obtained as (see e.g. Chapter 6 of \cite{Bha04})
 \begin{eqnarray}
\fg b &=& \frac{\zeta_2}{\sqrt{\Lambda_3-\Lambda_1}}\left(\sqrt{\Lambda_3(1-\Lambda_1)} \,\,\fg j_1+\xi\sqrt{\Lambda_1(\Lambda_3-1)}\,\,  \fg j_3\right), \quad\text{and}\nonumber\\
\tilde{\fg m} &=&\frac{\sqrt{\Lambda_3}-\sqrt{\Lambda_1}}{\zeta_2\sqrt{\Lambda_3-\Lambda_1}} \left(-\sqrt{{1-\Lambda_1} } \,\,\fg j_1+\xi\sqrt{\Lambda_3-1}\,\,  \fg j_3\right).
  \label{gleqsbcgg4kk}
 \end{eqnarray}
 The unit normal to the twin-twin boundary $\fg m$ is finally obtained using Eq. \eqref{gleqsbcgg4kk}$_2$ in Eq. \eqref{gleqsbcgg34b}$_2$
\begin{eqnarray}
{\fg m} =\frac{\sqrt{\Lambda_3}-\sqrt{\Lambda_1}}{\zeta_2\sqrt{\Lambda_3-\Lambda_1}} (\fg U_{tk} +\kappa_2 \fg n_2\otimes{\fg a}_2 )\cdot\left(-\sqrt{{1-\Lambda_1} } \,\,\fg j_1+\xi\sqrt{\Lambda_3-1}\,\,  \fg j_3\right),
  \label{gleqsbcgg4m}
 \end{eqnarray}
  where $\zeta_2$ in Eqs. \eqref{gleqsbcgg4kk} and \eqref{gleqsbcgg4m} is such that $|\fg m|=1$.
The rotation $\fg Q_3$ is then determined using  Eqs. \eqref{gleqsbcgg4kk}$_{1,2}$ in Eq. \eqref{gleqsbcgg32}.
 Note that the middle eigenvalues $\Lambda_2$ obtained would be an expression as a function of  the volume fractions $\kappa_1$ and $\kappa_2$. Setting $\Lambda_2= 1$, which is required for the existence of the twins within twins solution \cite{Chu_thesis1993},  would give a relation between $\kappa_1$ and $\kappa_2$.  However, it is not possible to obtain the exact solutions for  $\kappa_1$ and $\kappa_2$ from the limited governing equations at hand. The thickness of the transition layer $\kappa_t$ is also indeterminate.

 \subsection{Twins within twins solutions for cubic to tetragonal MTs}
 \label{solnTetrag}
The solutions for twins within twins for cubic to tetragonal MTs  are now obtained.  The three Bain stretch tensors for such transformations are  \cite{Bha04}
\begin{eqnarray}
\fg U_{t1} &=& \chi \,\fg c_1\otimes\fg c_1 +\alpha\, \fg c_2\otimes\fg c_2  + \alpha \,\fg c_3\otimes\fg c_3, \nonumber\\
\fg U_{t2} &=&\alpha\, \fg c_1\otimes\fg c_1 +\chi \, \fg c_2\otimes\fg c_2  + \alpha \,\fg c_3\otimes\fg c_3, \nonumber\\
\fg U_{t3} &=& \alpha \,\fg c_1\otimes\fg c_1 +\alpha\, \fg c_2\otimes\fg c_2  + \chi \, \,\fg c_3\otimes\fg c_3,
  \label{Bains}
 \end{eqnarray}
 where $\alpha<1$ and $\chi>1$ are the material constants and $\{\fg c_1, \,\fg c_2,\,\fg c_3\}$ is a right-handed standard Cartesian basis in the cubic unit cell of $\sf A$ such that the basis vectors are parallel to three mutually orthogonal sides of that unit cell.
The solutions for twins between $\sf M_1$-$\sf M_2$ are ($\fg n$ pointing into $\sf M_2$),  $\sf M_1$-$\sf M_3$  ($\fg n$ pointing into $\sf M_3$), and $\sf M_2$-$\sf M_3$  ($\fg n$ pointing into $\sf M_3$) are (see e.g. Chapter 5 of \cite{Bha04})
\begin{eqnarray}
\fg a &=&\sqrt{2}\upsilon(-\chi\fg c_1\pm\alpha\fg c_2), \qquad  \fg n =(\fg c_1\pm\fg c_2)/\sqrt{2}; \nonumber\\
\fg a &=&\sqrt{2}\upsilon(-\chi\fg c_1\pm\alpha\fg c_3), \qquad  \fg n =(\fg c_1\pm\fg c_3)/\sqrt{2};  \nonumber\\
\fg a &=&\sqrt{2}\upsilon(-\chi\fg c_2\pm\alpha\fg c_3), \qquad \fg n =(\fg c_2\pm\fg c_3)/\sqrt{2},
  \label{gleqsbcgg36}
 \end{eqnarray}
 respectively, where $\upsilon={(\chi^2-\alpha^2)}/({\chi^2+\alpha^2})$.
Since all the variants for tetragonal $\sf M$ phase are in  a twin relationship \cite{Bha04}, the combinations of possible twins within twins solutions   are $\{ \sf M_1,\sf M_2\}$-$\{\sf M_1,\sf M_3\}$,  $\{ \sf M_2,\sf M_1\}$-$\{\sf M_2,\sf M_3\}$, and $\{ \sf M_2,\sf M_3\}$-$\{\sf M_2,\sf M_1\}$, where the corresponding twin solutions $\{\fg a_1, \fg n_1\}$ and $\{\fg a_2, \fg n_2\}$ are to be considered from  Eq. \eqref{gleqsbcgg36}.

Using Eq. \eqref{gleqsbcgg34b}$_1$ and the solutions to the corresponding twin pairs from Eq.  \eqref{gleqsbcgg36},  we get ${\fg G}_2$ as the following diagonal tensor for all the possible combinations of twins within twins  listed above:
\begin{equation}
{\fg G}_2 =
\tilde{\fg A}^t\cdot\tilde{\fg A}
=\Lambda_1 \fg j_1\otimes \fg j_1 + \Lambda_2 \fg j_2\otimes \fg j_2 + \Lambda_3 \fg j_3\otimes \fg j_3, \quad \text{where}
  \label{gleqsbcgg34sgg}
 \end{equation}
\begin{eqnarray}
\Lambda_1 &=&\left[ 1-\kappa_2\upsilon\right]^2<1, \nonumber\\
\Lambda_2 &=&[(1+\kappa_1)\alpha^2+(1-\kappa_1)\chi^2]^2[(1+\kappa_2)\chi^2+(1-\kappa_2)\alpha^2]^2/(\chi^2+\alpha^2)^4=1, \quad\text{and}\quad\nonumber\\
\Lambda_3 &=& \left[ 1+\kappa_1\upsilon\right]^2>1,
  \label{gleqsbcgg34s}
 \end{eqnarray}
 and we have used the facts $\chi>1$, $\alpha<1$, $0<\upsilon<1$, and $0<\kappa_1,\kappa_2<1$. The eigenvectors $\fg j_1$, $\fg j_2$, and $\fg j_3$ are functions of the vectors $\fg c_1$, $\fg c_2$, and $\fg c_3$ and depend on the combinations of the variants in the twins as obtained below. We have imposed the condition $\Lambda_2=1$ in Eq. \eqref{gleqsbcgg34s}$_2$, which is a requirement for $\fg G_2$ given by Eq. \eqref{gleqsbcgg34sgg} to represent a twins within twins as discussed above, and  from that  the following two conditions on the parameters $\chi,\,\alpha,\,\kappa_1,\,\kappa_2$ are obtained:
 \begin{eqnarray}
 \frac{1}{\kappa_1}- \frac{1}{\kappa_2} &=&\upsilon, \quad \text{or} \nonumber\\
 (\kappa_1-\kappa_2)\frac{1}{\upsilon}+\kappa_1\kappa_2&=&\frac{2}{\upsilon^2}.
  \label{gleqsbcgg52}
 \end{eqnarray}
 Since $0<\upsilon<1$, Eq. \eqref{gleqsbcgg52}$_1$ is satisfied if and only if $\kappa_1<\kappa_2$. For $\kappa_1=\kappa_2$ Eq. \eqref{gleqsbcgg52}$_1$  yields the trivial condition $\chi=\alpha$ which  does not yield twins within twins. It is easy to verify that for all $\chi>1$ and $\alpha<1$ no $0<\kappa_1,\kappa_2<1$ satisfy Eq. \eqref{gleqsbcgg52}$_2$, and hence we disregard this relation. Finally, considering $\kappa_1$ and $\kappa_2$ are related by  Eq. \eqref{gleqsbcgg52}$_1$, and using the expressions for $\Lambda_1$ and $\Lambda_3$ given by Eqs. \eqref{gleqsbcgg34s}$_{1,3}$ into Eqs. \eqref{gleqsbcgg4} and \eqref{gleqsbcgg4m}, we get the solutions for $\fg b$ and $\fg m$  for different twins within twins as listed below.

\paragraph{Case I} For $\{ \sf M_1,\sf M_2\}$-$\{\sf M_1,\sf M_3\}$ twin pairs

For $\{ \sf M_1,\sf M_2\}$-$\{\sf M_1,\sf M_3\}$ twin pairs the indices are $i=k=1$, $j=2$, and $l=3$, and $\kappa_1$ and $\kappa_2$ are the volume fractions of $\sf M_1$ in the respective twins. The eigenvectors of $\fg G_2$ tensor are obtained as $\fg j_1=\fg c_3$, $\fg j_2=\fg c_1$,  and $\fg j_3=\fg c_2$. The vectors $\fg b$ and $\tilde{\fg m}$ are obtained using Eqs. \eqref{gleqsbcgg34s}$_{1,3}$ in Eq.  \eqref{gleqsbcgg4kk} as
 \begin{eqnarray}
 \fg b &=& \frac{\zeta_2}{\sqrt{(\kappa_1+\kappa_2)[2+(\kappa_1-\kappa_2)\upsilon]}}\left[\xi  (1-\kappa_2\upsilon) \sqrt{\kappa_1(2+\kappa_1\upsilon)}\,\fg c_2+(1+\kappa_1\upsilon) \sqrt{\kappa_2(2-\kappa_2\upsilon)}\,\fg c_3\right], \quad\text{and}\nonumber\\
 \tilde{\fg m}&=& \frac{\sqrt{\kappa_1+\kappa_2}\,\,\upsilon}{\zeta_2 \sqrt{2+(\kappa_1-\kappa_2)\upsilon}}\left[ \xi   \sqrt{\kappa_1(2+\kappa_1\upsilon)}\,\,\fg c_2- \sqrt{\kappa_2(2-\kappa_2\upsilon)}\,\,\fg c_3\right],
  \label{bmm}
 \end{eqnarray}
 respectively. Using Eq. \eqref{bmm}$_2$ and \eqref{gleqsbcgg4} in Eq. \eqref{gleqsbcgg4m} we finally get $\fg m$ as
  \begin{eqnarray}
 \fg m = \frac{\sqrt{\kappa_1+\kappa_2}\,\,\upsilon \alpha}{\zeta_2\sqrt{2+(\kappa_1-\kappa_2)\upsilon}}\left[- \upsilon\kappa_2^{1.5}\sqrt{2-\kappa_2\upsilon}\,\,\fg c_1 +\xi  \sqrt{\kappa_1(2+\kappa_1\upsilon)}\,\,\fg c_2+  \sqrt{\kappa_2(2-\kappa_2\upsilon)}(1+\kappa_2\upsilon)\,\,\fg c_3\right],
  \label{bmmm}
 \end{eqnarray}
 where the condition $|\fg m|=1$ yields
 \begin{eqnarray}
 \zeta_2= \frac{\sqrt{\kappa_1+\kappa_2}\,\,\upsilon \alpha}{\sqrt{2+(\kappa_1-\kappa_2)\upsilon}}\left[\kappa_2(2-\kappa_2\upsilon)(1+2\kappa_2\upsilon+2\kappa_2^2\upsilon^2)+\kappa_1(2+\kappa_1\upsilon)\right].
  \label{zeta}
 \end{eqnarray}

\paragraph{Case II} For $\{ \sf M_2,\sf M_1\}$-$\{\sf M_2,\sf M_3\}$ twin pairs

 In this case the indices are $i=k=2$, $j=1$, and $l=3$, and $\kappa_1$ and $\kappa_2$ are the volume fractions of $\sf M_2$ in the respective twins. The eigenvectors for $\fg G_2$ tensor are given by $\fg j_1=\fg c_3$, $\fg j_2=\fg c_2$,  and $\fg j_3=\fg c_1$. The vectors $\fg b$ and $\tilde{\fg m}$ are obtained using Eqs. \eqref{gleqsbcgg4} and \eqref{gleqsbcgg4kk} as
  \begin{eqnarray}
 \fg b &=& \frac{\zeta_2}{\sqrt{(\kappa_1+\kappa_2)[2+(\kappa_1-\kappa_2)\upsilon]}}\left[(1+\kappa_1\upsilon) \sqrt{\kappa_2(2-\kappa_2\upsilon)}\,\fg c_3 +\xi  (1-\kappa_2\upsilon) \sqrt{\kappa_1(2+\kappa_1\upsilon)}\,\fg c_1\right], \quad\text{and}\nonumber\\
 \tilde{\fg m}&=& \frac{\sqrt{\kappa_1+\kappa_2}\,\,\upsilon}{\zeta_2 \sqrt{2+(\kappa_1-\kappa_2)\upsilon}}\left[- \sqrt{\kappa_2(2-\kappa_2\upsilon)}\,\,\fg c_3 +\xi   \sqrt{\kappa_1(2+\kappa_1\upsilon)}\,\,\fg c_1\right],
  \label{bmm1}
 \end{eqnarray}
 respectively. Using Eq. \eqref{bmm}$_2$ in Eq. \eqref{gleqsbcgg4m} we finally get $\fg m$ as
  \begin{eqnarray}
 \fg m = \frac{\sqrt{\kappa_1+\kappa_2}\,\,\upsilon \alpha}{\zeta_2\sqrt{2+(\kappa_1-\kappa_2)\upsilon}}\left[- \xi\upsilon\kappa_2^{1.5}\sqrt{2-\kappa_2\upsilon}\,\,\fg c_2 +\xi  \sqrt{\kappa_1(2+\kappa_1\upsilon)}\,\,\fg c_1+  \sqrt{\kappa_2(2-\kappa_2\upsilon)}(1+\kappa_2\upsilon)\,\,\fg c_3\right],
  \label{bmmm1}
 \end{eqnarray}
 where $\zeta_2$ is given by Eq. \eqref{zeta}.

\paragraph{Case III} For $\{ \sf M_3,\sf M_1\}$-$\{\sf M_3,\sf M_2\}$ twin pairs

In this case the indices are $i=k=3$, $j=1$, and $l=2$, and $\kappa_1$ and $\kappa_2$ are the volume fractions of $\sf M_3$ in the respective twins. The eigenvectors for $\fg G_2$ tensor are given by $\fg j_1=\fg c_2$, $\fg j_2=\fg c_3$,  and $\fg j_3=\fg c_1$. The vectors $\fg b$ and $\tilde{\fg m}$ are obtained using Eqs. \eqref{gleqsbcgg4} and \eqref{gleqsbcgg4m} as
 \begin{eqnarray}
 \fg b &=& \frac{\zeta_2}{\sqrt{(\kappa_1+\kappa_2)[2+(\kappa_1-\kappa_2)\upsilon]}}\left[(1+\kappa_1\upsilon) \sqrt{\kappa_2(2-\kappa_2\upsilon)}\,\fg c_2 +\xi  (1-\kappa_2\upsilon) \sqrt{\kappa_1(2+\kappa_1\upsilon)}\,\fg c_1\right], \quad\text{and}\nonumber\\
 \tilde{\fg m}&=& \frac{\sqrt{\kappa_1+\kappa_2}\,\,\upsilon}{\zeta_2 \sqrt{2+(\kappa_1-\kappa_2)\upsilon}}\left[- \sqrt{\kappa_2(2-\kappa_2\upsilon)}\,\,\fg c_2 +\xi   \sqrt{\kappa_1(2+\kappa_1\upsilon)}\,\,\fg c_1\right],
  \label{bmm2}
 \end{eqnarray}
 respectively. Using Eq. \eqref{bmm}$_2$ in Eq. \eqref{gleqsbcgg4m} we finally get $\fg m$ as
  \begin{eqnarray}
 \fg m = \frac{\sqrt{\kappa_1+\kappa_2}\,\,\upsilon \alpha}{\zeta_2\sqrt{2+(\kappa_1-\kappa_2)\upsilon}}\left[- \xi\upsilon\kappa_2^{1.5}\sqrt{2-\kappa_2\upsilon}\,\,\fg c_3 +\xi  \sqrt{\kappa_1(2+\kappa_1\upsilon)}\,\,\fg c_1+  \sqrt{\kappa_2(2-\kappa_2\upsilon)}(1+\kappa_2\upsilon)\,\,\fg c_2\right],
  \label{bmmm2}
 \end{eqnarray}
  where $\zeta_2$ is given by Eq. \eqref{zeta}.

 In summary, we have obtained the general analytical solution for  $\fg a_1$, $\fg a_2$, $\fg n_1$, $\fg n_2$, $\fg b$,  and $\fg m$ listed in Eqs. \eqref{gleqsbcgg4}, \eqref{gleqsbcgg4kk}, and \eqref{gleqsbcgg4m}. The rotation tensors $\fg Q_1$, $\fg Q_2$, and $\fg Q_3$ can be finally obtained using Eqs. \eqref{gleqsbcgg31}$_{1,2}$ and \eqref{gleqsbcgg2}. The twins within twins solutions for cubic to tetragonal MTs are listed in Eqs. \eqref{gleqsbcgg36} and \eqref{bmm} to \eqref{bmmm2}. Since the volume fractions $\kappa_1$ and $\kappa_2$ satisfy the relation given by Eq. \eqref{gleqsbcgg52}$_1$, there are obviously many solutions possible for each of the twins within twins listed in Cases (I), (II) and (III) (also see Chapter 7 of \cite{Bha04}). The width of the twins within twins interface (shaded region in Fig. \ref{twin_twin}) $\delta_\kappa$ is indeterminate within the set of governing equations at hand.

\section{Results and discussions}
\label{results}
We now present the simulation results for twins within twins microstructure obtained using our phase-field approach. The materials properties for  NiAl alloy,  which exhibits cubic to tetragonal MTs,  are considered and listed in Section \ref{propss}. In Section \ref{pfresultss} the phase-field results are compared with the crystallographic solution obtained in Section \ref{crystTh}.

\subsection{ Material parameters}
\label{propss}
\begin{figure}[t!]
\centering
  \includegraphics[width=2.8in, height=2.4in] {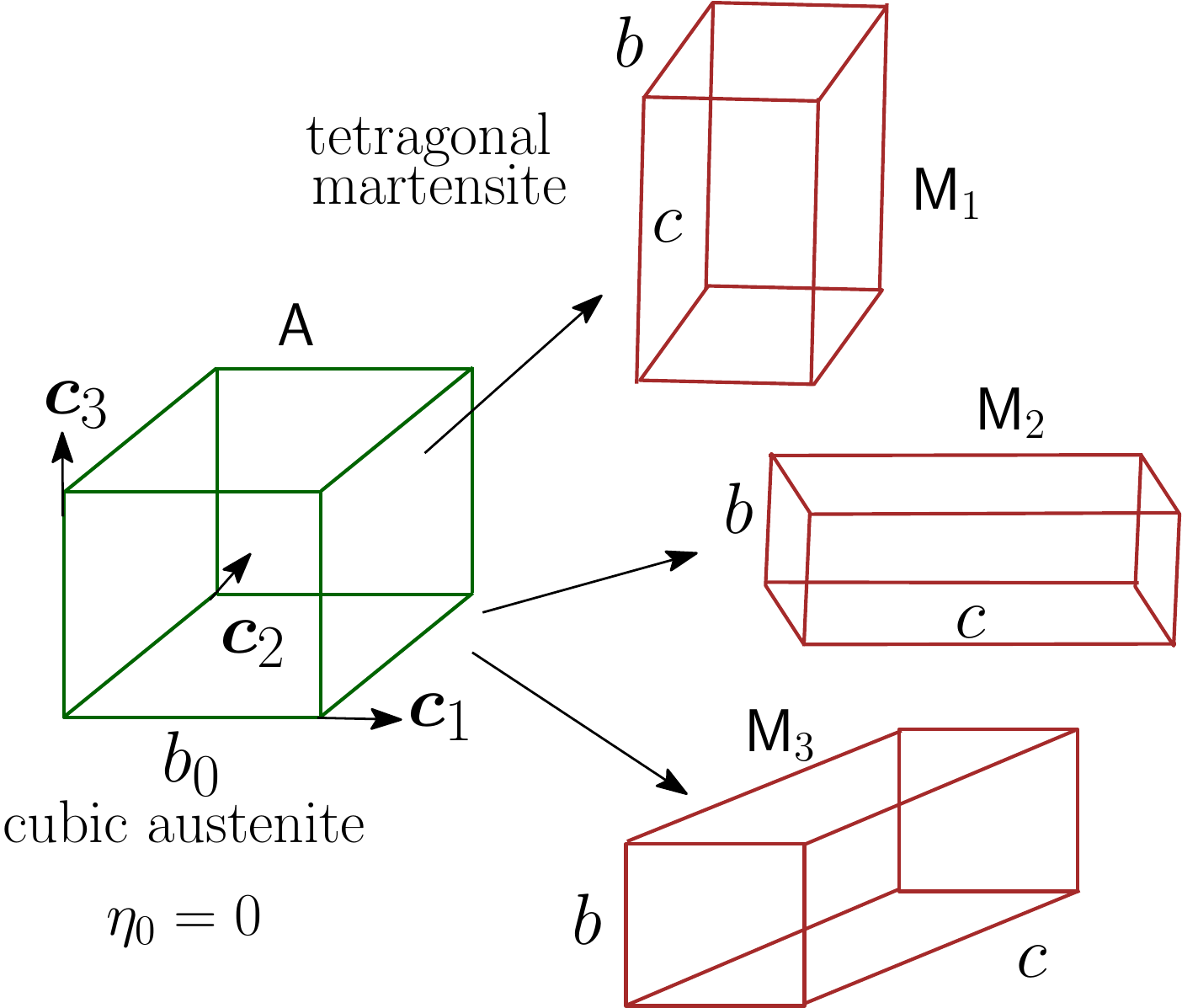}
\caption{Unit cells for cubic austenite and three tetragonal martensitic variants.}
\label{unit_cells1}
\end{figure}
The material parameters for NiAl alloy are enlisted here. We consider the interfacial widths and energies as  $\delta_{0M}=1$ nm, $\gamma_{0M}=0.2$ N/m, $\delta_{12}= \delta_{13}=\delta_{23}=0.75$ nm, and $\gamma_{12}=\gamma_{13}=\gamma_{23}=0.1$ N/m. Using the following analytical relations between the interfacial thickness and energy, and the phase-field parameters \cite{Levitas-14a,Steinbach-09}
\begin{equation}
\delta_{ij}=\sqrt{18\beta_{ij}/A_{ij}};  \qquad \beta_{ij}=\gamma_{ij}\delta_{ij} \quad \text{for } \sf A\text{-}\sf M \text{  and all the variant-variant interfaces},
\label{analyticalsols}
\end{equation}
which were obtained by solving an 1D Ginzburg-Landau equation neglecting mechanics, we obtain $\rho_0A_{0M} =3600$ MPa, $\rho_0\bar{A}=2400$ MPa, $\beta_{0M}= 2\times 10^{-10}$ N, and $\beta_{12}=\beta_{13}=\beta_{23}=7.5\times 10^{-11}$ N. We take $\theta_e=215$ K and $\Delta s = -1.47$ MPa K$^{-1}$, using which we calculate  the critical temperatures for ${\sf A}\to{\sf M}$ and ${\sf M}\to{\sf A}$  transformations as (see \cite{Levitas-Roy-Acta-16}) $\theta^c_{{\sf A}\to{\sf M}}= \theta_e+A_{0M}/(3\Delta s) =0$ K and  $\theta^c_{{\sf M}\to{\sf A}}= \theta_e-A_{0M}/(3\Delta s) =430$ K, respectively. The  Lam\'{e} constants assuming isotropic elasticity are taken  to be identical for all the phases $\sf A$, ${\sf M}_1$, ${\sf M}_2$, and ${\sf M}_3$:  $\lambda_0=\lambda_1=\lambda_2 =\lambda_3=74.62 $ GPa, $\mu_0=\mu_1=\mu_2=\mu_3=72$ GPa . The other material constants are taken as $a_\beta=a_\varepsilon=a_K=3$, $a_0=10^{-3}$, $K_{12}= K_{23}=K_{13}=50 $ GPa, $K_{012}= K_{023}=K_{013}=5 $ GPa, $K_{0123}= K_{123}=50 $ GPa,   $L_{0M}=2600$ (Pa-s)$^{-1}$ and $L_{12}=L_{13}=L_{23}=12600$ (Pa-s)$^{-1}$.  The transformation stretches  are $\alpha = 0.922$ and $\chi=1.215$ \cite{Levitas-preston-PRB-I}.

\subsection{Numerical result for twins within twins}
\label{pfresultss}
\begin{figure}[t!]
\centering
  \includegraphics[width=4.0in, height=3.5in] {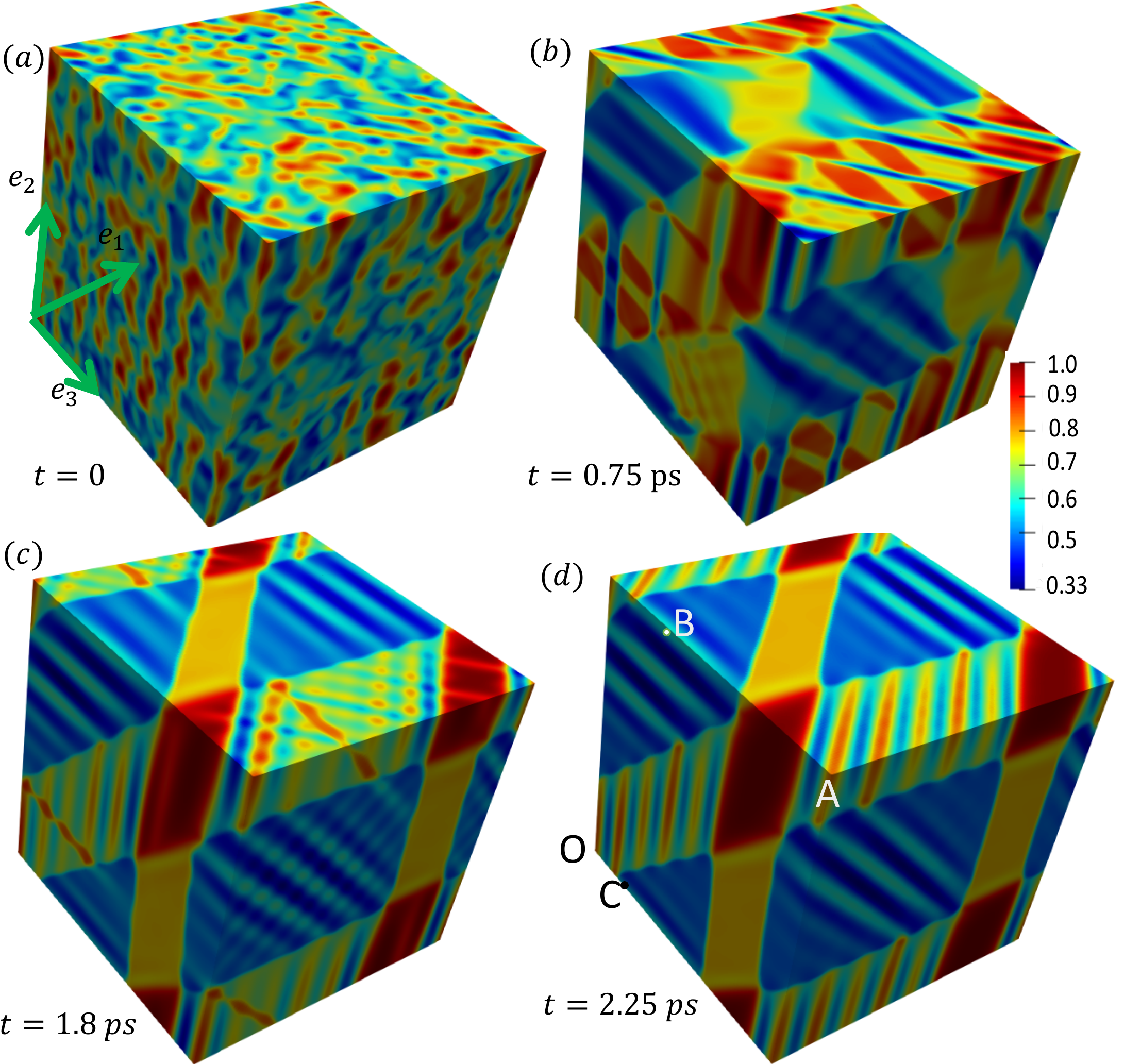}
\caption{Evolution of twin within twin microstructure in a $22$ nm$\times 22$ nm$\times 22$ nm cube shown by color plot of $\eta_{eq}=\eta_0(1-0.67\eta_1-0.33\eta_2)$: $\eta_{eq}=0.33$ denotes $\sf M_1$; $\eta_{eq}=0.67$ denotes $\sf M_2$; $\eta_{eq}=1$ denotes $\sf M_3$.}
\label{Eta_eq}
\end{figure}

\begin{figure}[t!]
\centering
  \includegraphics[width=4.0in, height=1.7in] {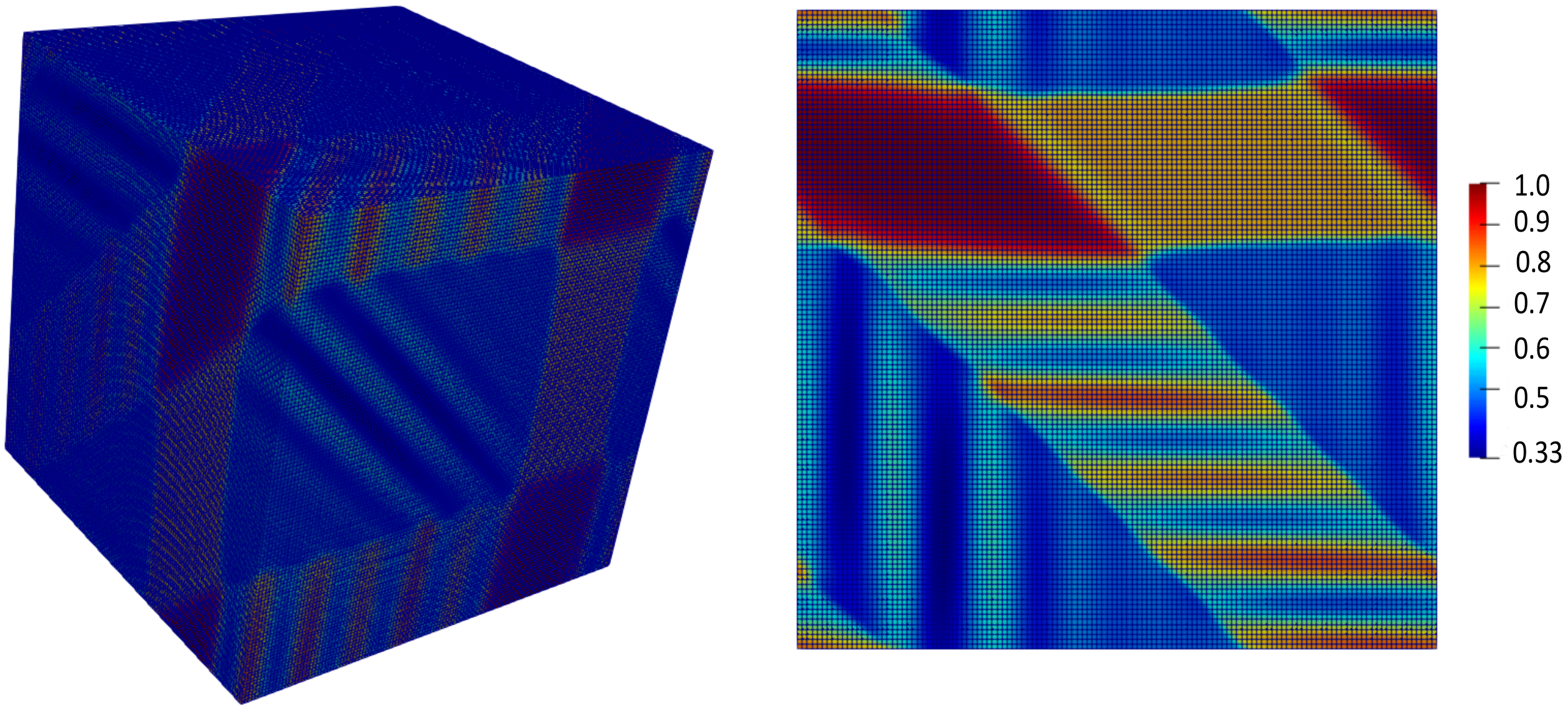}
\caption{(a) Mesh density in the 3D computational domain. (b) Mesh density on one of the external boundaries. }
\label{MeshDensity}
\end{figure}

For the  simulation we consider a $22$ nm$\times 22$ nm$\times 22$ nm cube as the reference configuration $\Omega_0$ as shown in Figs. \ref{Eta_eq}(a).  The periodic BCs  given by Eq. \eqref{PBC1} are used for all the order parameters on the respective opposite faces of the cube domain. We have used the periodic BC for the normal component of the displacement vector on the opposite faces of the cube given by Eq. \eqref{PBC2} with the homogeneous deformation gradient as $\fg F_h =\fg I+ 0.98\,\fg n_0\otimes \fg n_0$, where $\fg n_0$ is the unit normal to the opposite faces of the cube in $\Omega_0$. On each face of the cube domain, we have used the traction-free BC for the tangential components of the first Piola-Kirchhoff traction vector.  The temperature of the sample is taken as $\theta=0$ K. The material properties listed in Section \ref{propss} are used.   The Bain tensors listed in Eq. \eqref{Bains} are used in Eq. \eqref{utilde}, where the  basis vectors $\fg c_1$, $\fg c_2$, and $\fg c_3$ of $\sf A$ unit cell are parallel to basis vectors $\fg e_1$, $\fg e_2$, and $\fg e_3$, respectively, attached to the  sample (see Fig. \ref{Eta_eq} (a)). The initial distribution of the order parameters are taken between $0\leq \eta_0\leq 1$, $0\leq \eta_1\leq 0.8$, and $0\leq \eta_2\leq 0.8$, all distributed randomly shown in Figs. \ref{Eta_eq}(a). The other order parameter $\eta_3$ is calculated using Eq. \eqref{constraintss} for all the times $t\geq 0$. We have used a finite element procedure similar to that of \cite{Basak-Levitas-2019CMAME} developed by the authors. A finite element code has been developed using an open-source code deal.ii \cite{Bangerth-16}. The domain is discretized  spatially with quadratic brick element and it is ensured that at least three grid points lie across all the interfaces. The mesh density in the 3D domain is shown in Fig. \ref{MeshDensity}(a), and the mesh density on one of the boundaries is shown in Fig. \ref{MeshDensity}(b).
The time derivatives of the order parameters are discretized using the backward difference scheme of order two described in \cite{Basak-Levitas-2019CMAME}. A constant time step size of $2\times 10^{-16}$ s is used for the simulation.

\begin{figure}[t!]
\centering
  \includegraphics[width=4.0in, height=1.7in] {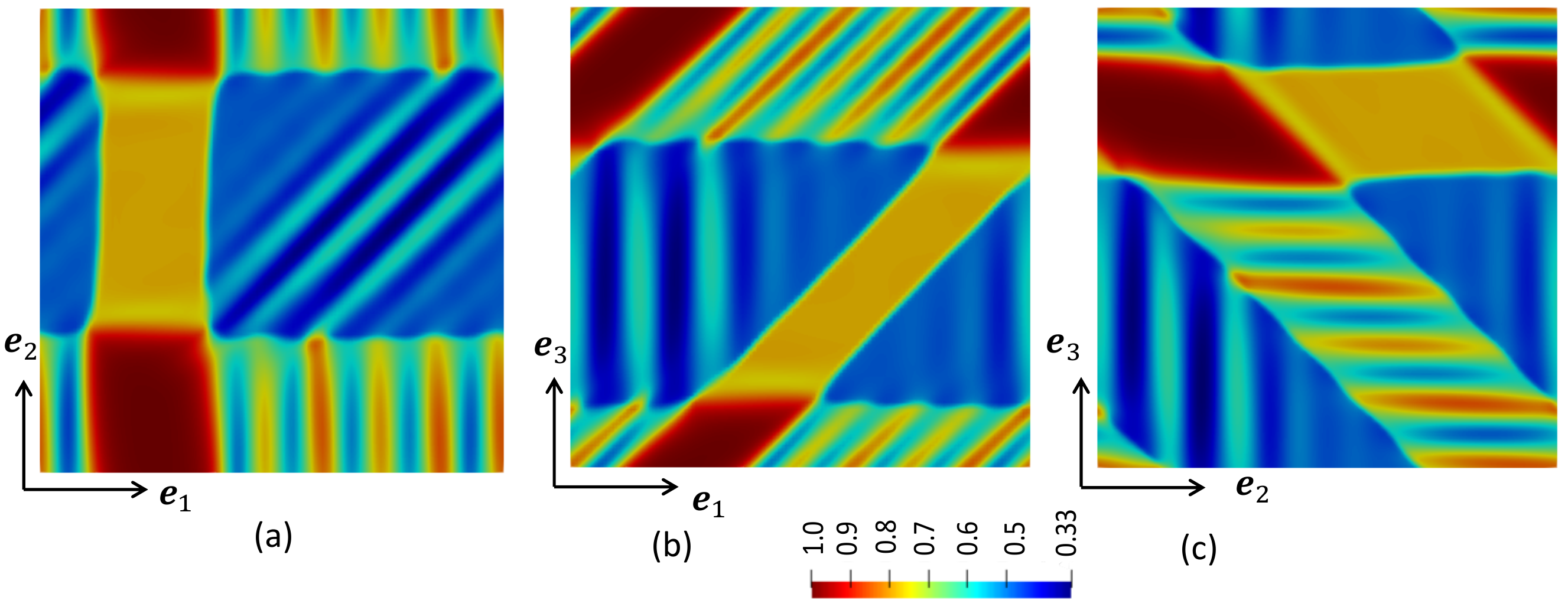}
\caption{The twin microstructures on three mutually perpendicular faces of the sample shown in Fig. \ref{Eta_eq}. The colour plots of $\eta_{eq}$ are shown: $\eta_{eq}=0.33$ denotes $\sf M_1$; $\eta_{eq}=0.67$ denotes $\sf M_2$; $\eta_{eq}=1$ denotes $\sf M_3$.  }
\label{Eta_eq_2D}
\end{figure}

 The evolution of the  microstructure is shown  at different time instances in Figs. \ref{Eta_eq}. We have in fact shown a colour plot for an equivalent order parameter defined as $\eta_{eq}=\eta_0(1-0.67\eta_1-0.33\eta_2)$ which obviously takes the following values in different phases: $\eta_{eq}=0$ in $\sf A$, $\eta_{eq}=0.33$ in $\sf M_1$, $\eta_{eq}=0.67$ in $\sf M_2$, and $\eta_{eq}=1$ in $\sf M_3$.  Figure  \ref{Eta_eq}(a) shows the initial distribution of $\eta_{eq}$. Figures  \ref{Eta_eq}(b) and (c) show the intermediate microstructures at different time instances which are approaching  a twinned microstructure. We finally obtain  twins within twins microstructures between the twin pairs $\sf M_1$-$\sf M_2$ and $\sf M_1$-$\sf M_3$ as shown in Fig.   \ref{Eta_eq}(d).
 The microstructure shown in Fig.   \ref{Eta_eq}(d) is a little far from being a stationary one. However, the twinned microstructure obtained can be compared with the analytical solution as there is no further significant change in the orientations of the twin boundaries and twin-twin boundaries with time as observed  in Figs. \ref{Eta_eq} (c) and (d). The microstructures on three faces of the domain with unit normals parallel to $\fg e_3$, $\fg e_2$ and $\fg e_1$ are shown in Fig. \ref{Eta_eq_2D}(a), (b), and (c), respectively.

\begin{figure}[t!]
\centering
  \includegraphics[width=5.0in, height=3.5in] {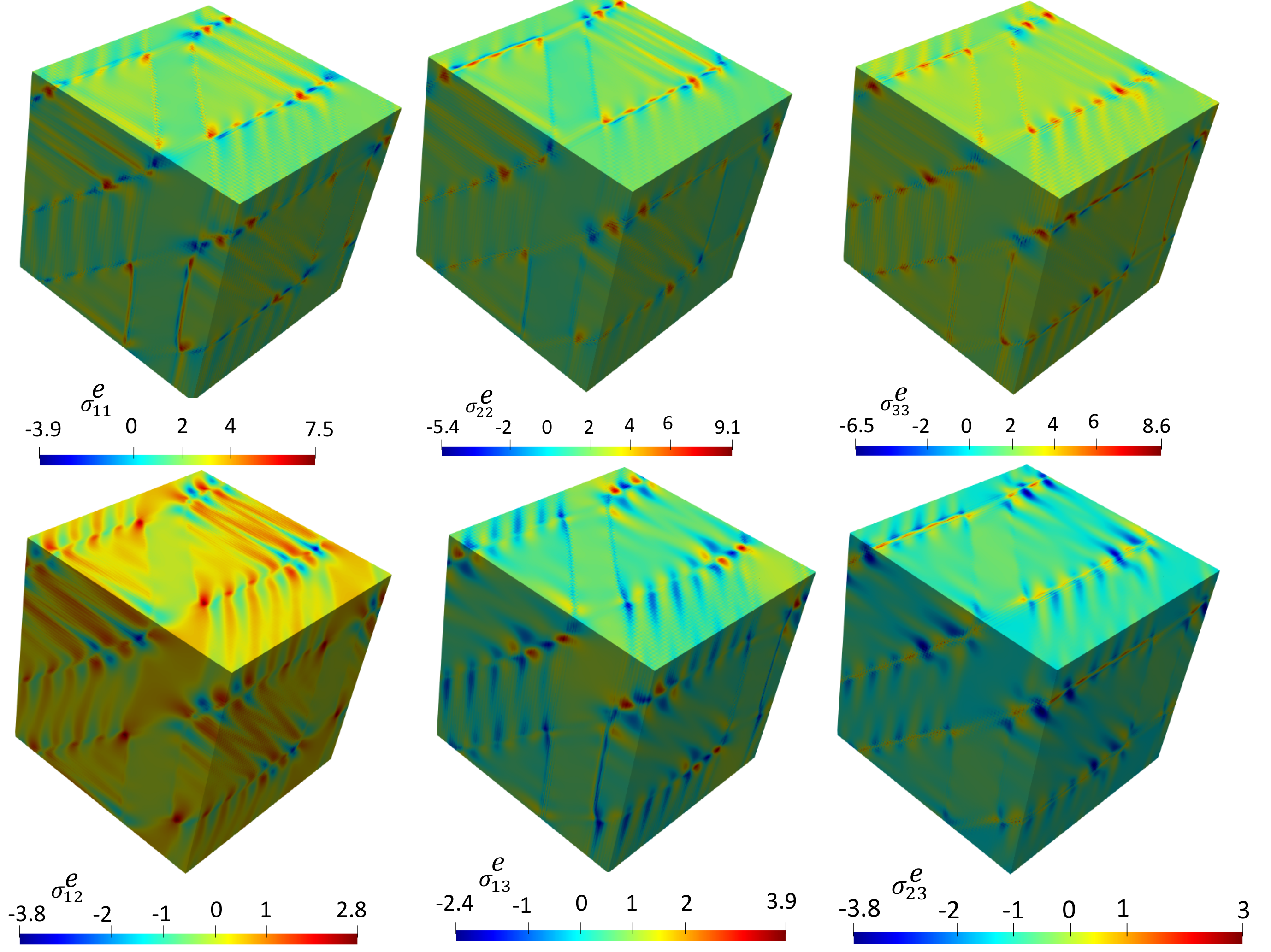}
\caption{Components of the Cauchy elastic stress tensor (in GPa) in the domain shown in Fig.  \ref{Eta_eq}(d). }
\label{3Dstress}
\end{figure}

\begin{figure}[t!]
\centering
\hspace{-8mm}
\subfigure[]{
  \includegraphics[width=3.1in, height=2.8in] {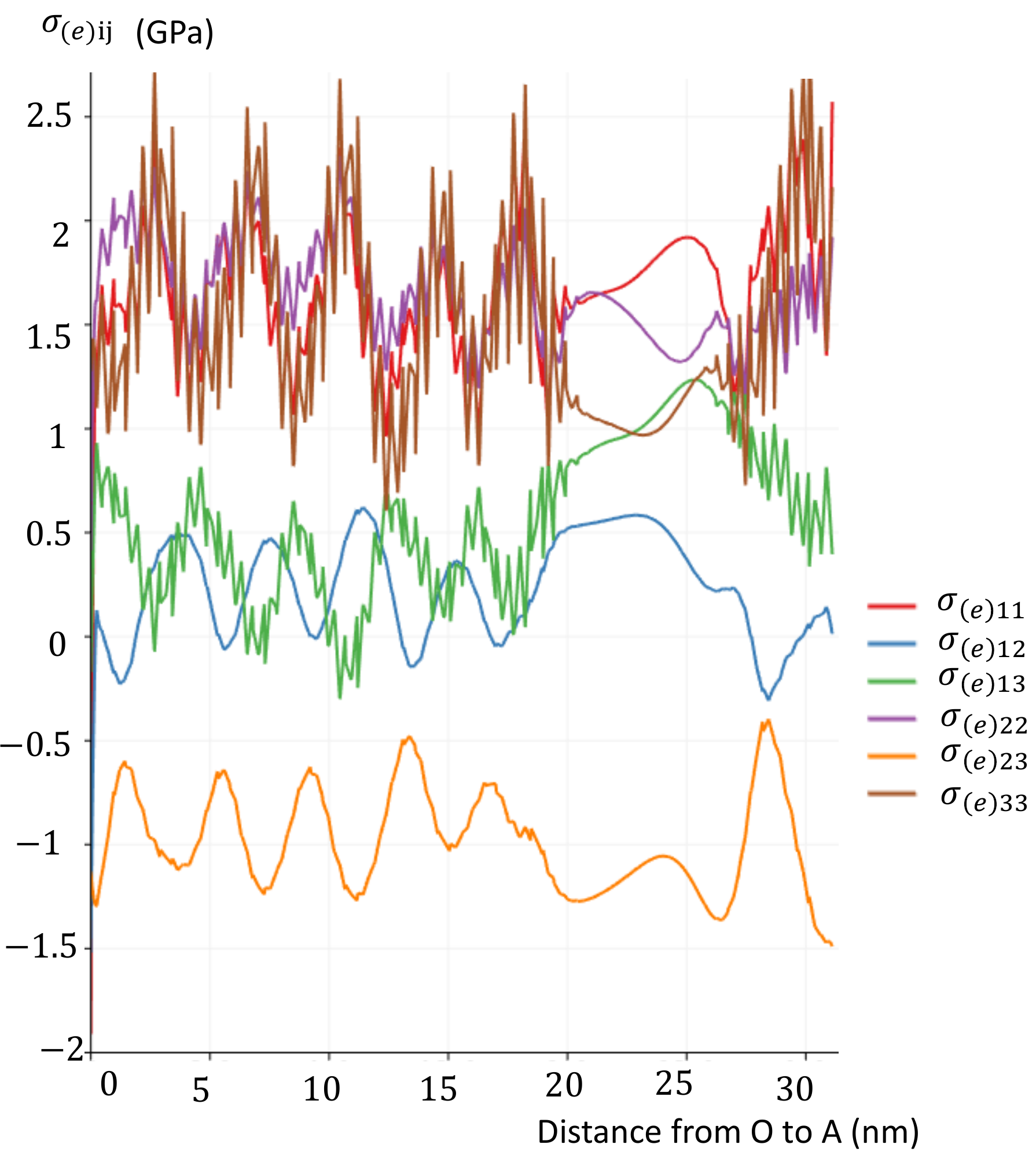}
	}\hspace{-3mm}
    \subfigure[]{
    \includegraphics[width=2.7in, height=2.8in] {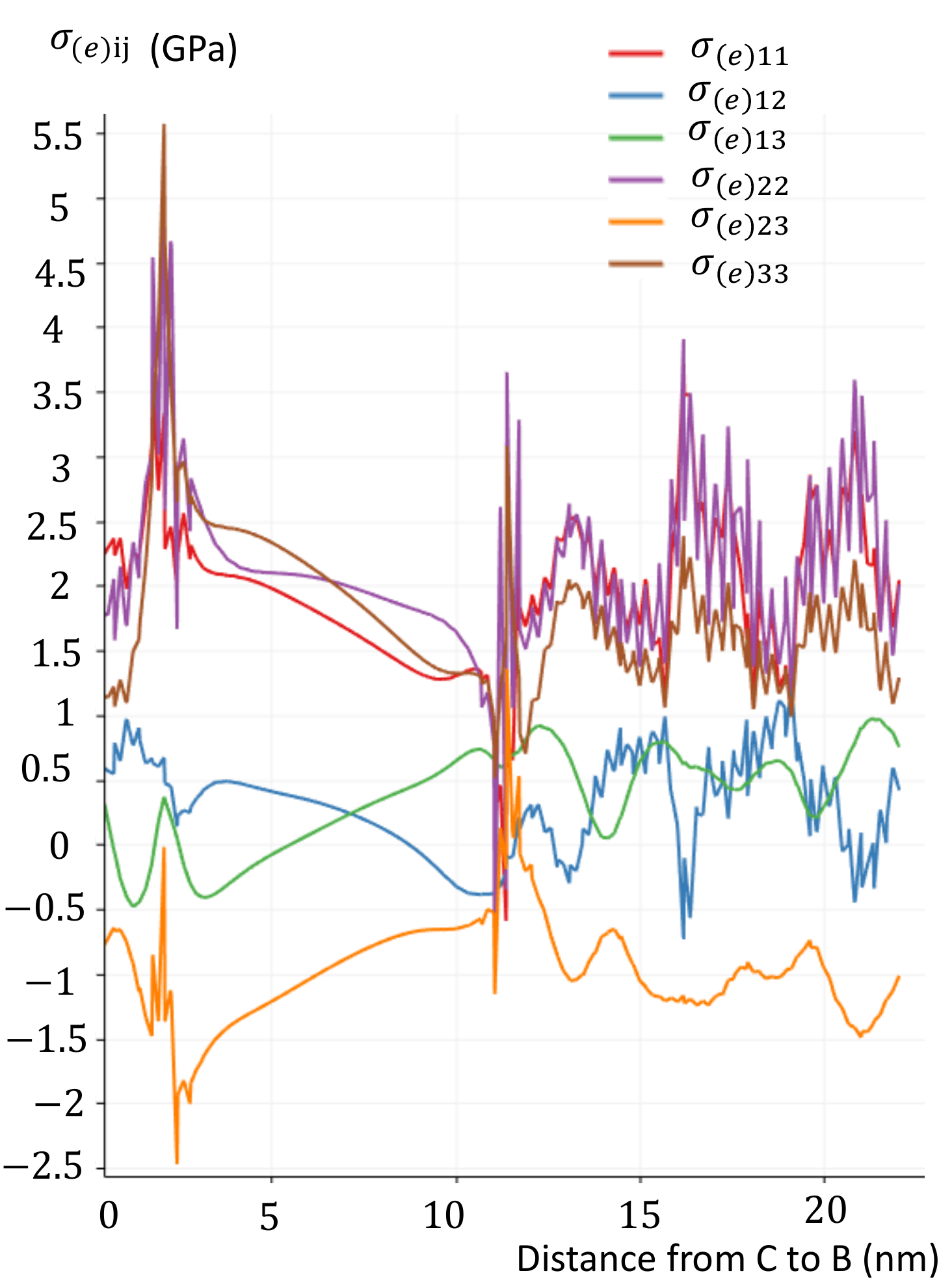}
}
\caption{ Components of the Cauchy elastic stress tensor across the lines  drawn between points (a) O and A, and (b) C and B  shown in Fig.  \ref{Eta_eq}(d). Obviously, figures (a) and (b) show the elastic stress distribution across the twin boundaries between variants $\sf M_1$-$\sf M_3$ and $\sf M_1$-$\sf M_2$, respectively.}
\label{interf_stresss}
\end{figure}

 The plots for the components of the Cauchy elastic stress tensor (in GPa) at $t=2.25$ ps (corresponding to the microstructure shown in Fig. \ref{Eta_eq} (d)) are shown in Fig. \ref{3Dstress}. The internal stresses are concentrated mainly across the twin-twin boundaries and also across the twin boundaries (see e.g. \cite{Schuh-13} for experimental results). The stresses vary from compressive to tensile between two adjacent variant plates within and near the twin-twin interfaces.
 For a better understanding of the elastic stresses across the interfaces, we have shown a plot for the  components across   the lines joined by the points O and A, and  C and B (see Fig. \ref{Eta_eq} (d)) in Fig. \ref{interf_stresss}(a) and (b), respectively.  Obviously, in these two figures, the elastic stresses within $\sf M_1$-$\sf M_3$ and $\sf M_1$-$\sf M_2$ twin boundaries are plotted. Figure \ref{interf_stresss}(b) also shows the stresses across a twin-twin boundary.
 All the normal stresses $\sigma_{(e)11}$, $\sigma_{(e)22}$, and $\sigma_{(e)33}$ are significantly higher across the twin boundaries compared to the adjacent phases. The reason for such large elastic stresses within the twin boundaries is studied in detail by the authors in \cite{Basak-Levitas-2017-ActaMater}. The shear stresses $\sigma_{(e)12}$ and $\sigma_{(e)13}$ on the corresponding external boundary (having unit normal $\fg e_1$ in $\Omega_0$) are much lower due to the traction-free BC applied in the tangential plane. From the elastic stress plots shown in Fig. \ref{3Dstress} it is clear that stresses are mainly concentrated within the twin boundaries and the twins-twins boundaries. In fact, the stresses within   twins within twins boundaries are usually much higher than that across the twin boundaries (also see Fig. \ref{interf_stresss}).
 This can be justified by noticing that the twin-twin boundaries are compatible in an average sense, whereas the twin boundaries are compatible in Hadamard's sense according to the crystallographic theory (see Chapter 5 of \cite{Bha04}). Understanding the stress distributions across these interfaces is important from the materials design perspective \cite{Schuh-13}.

 \begin{table}
\centering
Table-1: Crystallographic solutions for twins between the variants for NiAl alloy in $\{\fg c_1\,,\fg c_2\,,\fg c_3  \}$ basis.
\begin{tabular}{ |p{2.2cm}|p{4.2cm}|p{4.2cm}|}
 \hline
  Variant  pair &  $\fg a$ &  $\fg n$  \\
 \hline
 $\sf M_1$-$\sf M_2$&   $ -0.4625\,\fg c_1\pm 0.3510\,\fg c_2$ &   $0.7071\,\fg c_1\pm 0.7071\,\fg c_2$   \\
 \hline
 $\sf M_2$-$\sf M_3$& $0.4625\,\fg c_1\pm 0.3510\,\fg c_3$&   $-0.7071\,\fg c_1\pm 0.7071\,\fg c_3$   \\
 \hline
 $\sf M_3$-$\sf M_1$ &   $0.4625\,\fg c_2\pm 0.3510\,\fg c_3 $ &   $ -0.7071\,\fg c_2\pm 0.7071\,\fg c_3$  \\
 \hline
\end{tabular}
\label{table1}
\end{table}
 \begin{figure}[t!]
\centering
  \includegraphics[width=2.6in, height=2.2in] {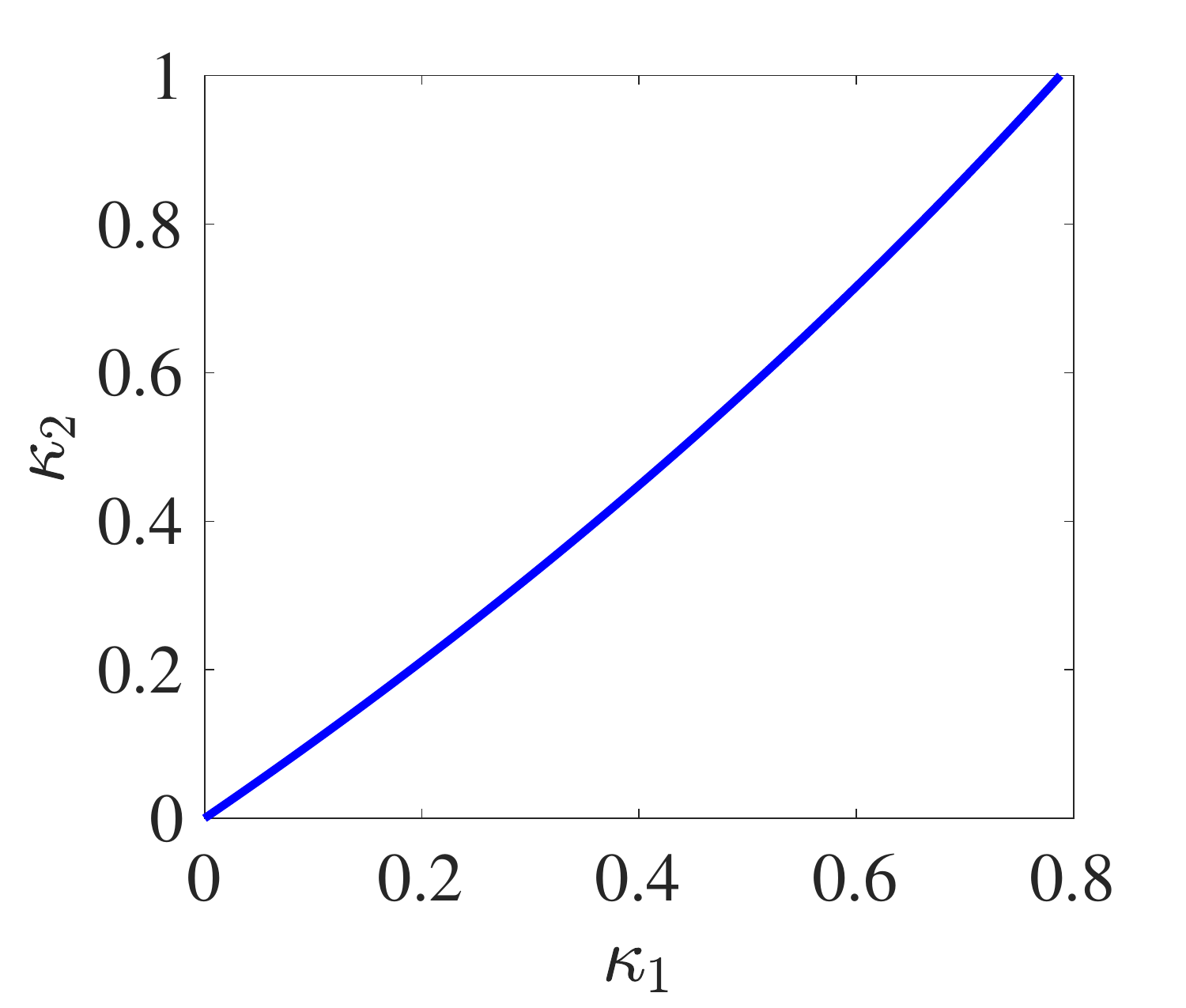}
\caption{Plot for $\kappa_2$ versus $\kappa_1$ given by Eq. \eqref{gleqsbcgg52}$_1$ for NiAl alloy.}
\label{unit_cells}
\end{figure}

\begin{figure}[t!]
\centering
  \includegraphics[width=2.8in, height=2.8in] {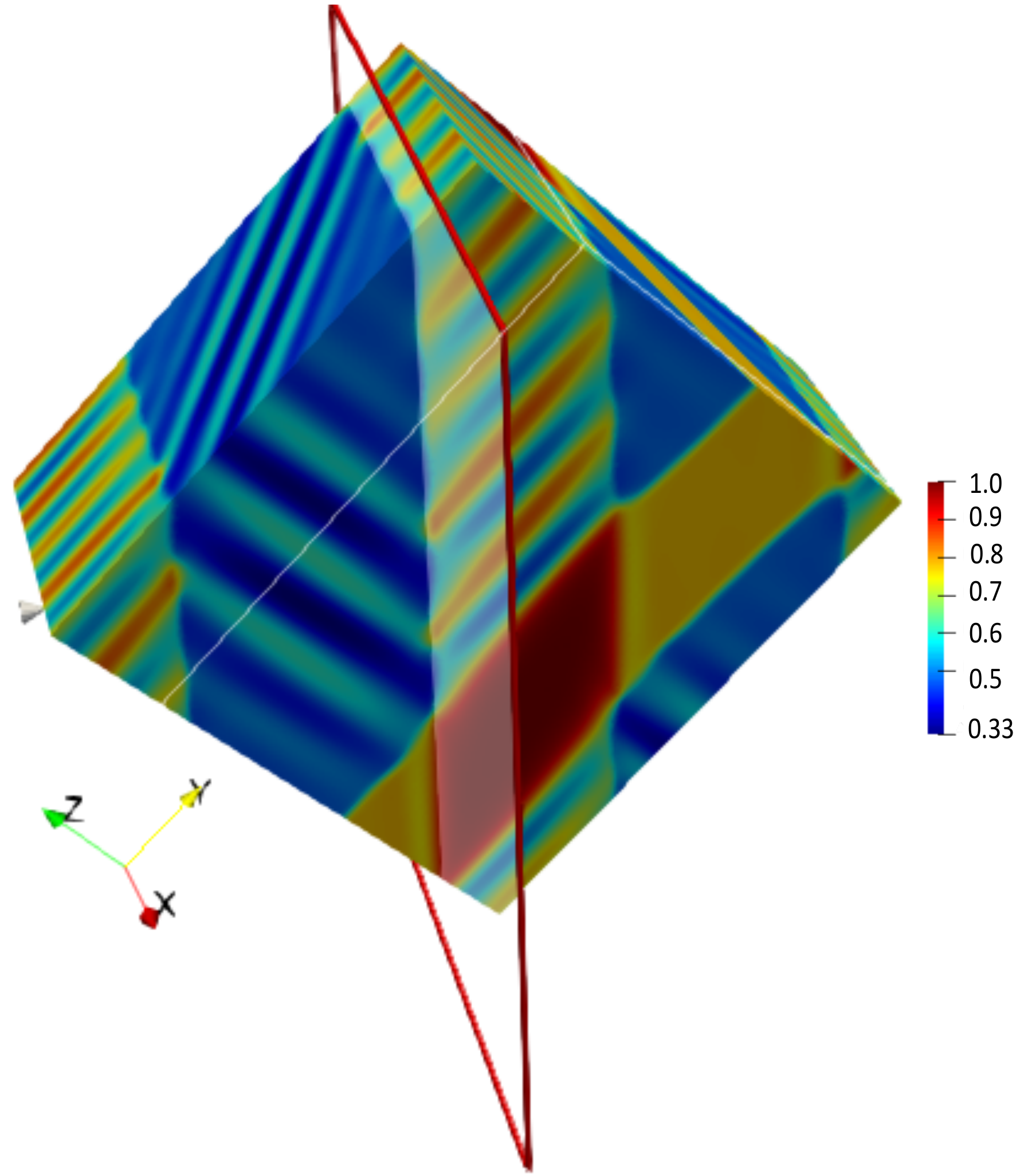}
\caption{A plane within a twins within twins interface (within the domain shown in Fig. \ref{Eta_eq}(d)) is shown by the rectangle. The unit normal to that plane is $\fg m =0.1020\,\fg c_1-0.7204\,\fg c_2+0.6859\,\fg c_3$.}
\label{twin_plane}
\end{figure}
\paragraph{Comparison of crystallographic and numerical solutions. }
We now present a comparative study of the microstructure obtained numerically with the crystallographic solution given in Section \ref{solnTetrag}. We have shown in Fig. \ref{Eta_eq} the twins within twins microstructures obtained using the present phase-field approach, where  a twin with a variant pair ${\sf M}_1$-${\sf M}_2$ is forming interfaces with another twin made of variants pair ${\sf M}_1$-${\sf M}_3$, i.e. the indices of Fig. \ref{twin_twin} are $i=k=1$, $j=2$, and $l=3$. The normal to the interfaces between ${\sf M}_1$ and ${\sf M}_2$ plates are making approximately  $45^\circ$ with both $\fg e_1$ and $\fg e_2$ axes, and the normal to the interfaces between ${\sf M}_1$ and ${\sf M}_3$ plates are making approximately  $45^\circ$ with both $\fg e_2$ and $\fg e_3$ axes. The normals to these twins are obviously in good agreement with the crystallographic solution listed in Table-1. The volume fractions of ${\sf M}_1$ in the respective twin pairs are calculated as $\kappa_1=0.46$ and $\kappa_2=0.56$.  The unit normal to one of  the twins within twins interface  $\fg m$ is obtained as
 \begin{equation}
\fg m =0.1020\,\fg c_1-0.7204\,\fg c_2+0.6859\,\fg c_3,
  \label{gleqsbcgg41}
 \end{equation}
 where a rectangular plane (on red lines) lying on the twin-twin interface is shown in  Fig. \ref {twin_plane}.
We now calculate the analytical expression for $\fg m$.  As mentioned earlier, the volume fractions $\kappa_1$ and $\kappa_2$,  satisfying the relation given by Eq. \eqref{gleqsbcgg52}$_1$ and plotted in Fig. \ref{unit_cells},  cannot be uniquely determined from the limited  governing equations at hand. In  view of that, we simply assume $\kappa_1=0.46$ for the analytical solution and obtain $\kappa_2=0.5250$ using Eq. \eqref{gleqsbcgg52}$_1$ which differs from the numerical result by $7.7$\%. Finally, using Eqs. \eqref{bmm}, \eqref{bmmm}, and  \eqref{zeta} we obtain $\zeta_2=0.2634$ and the vectors $\fg b$, $\tilde{\fg m}$, and $\fg m$ as
\begin{eqnarray}
\fg b &=& \pm 0.1362\,\fg c_2+0.2093\,\fg c_3, \nonumber\\
 \tilde{\fg m} &=& \pm 0.7119\,\fg c_2-0.7115\,\fg c_3, \quad \text{and}\nonumber\\
\fg m &=& - 0.0928\,\fg c_1\pm 0.6560\,\fg c_2+0.7492\,\fg c_3.
  \label{gleqsbcgg41m}
 \end{eqnarray}
 The maximum difference in the components of analytical and numerically obtained unit normals $\fg m$ is within 10\%, and the orientation of the analytical $\fg m$ differs by $12.3^\circ$ from the numerical one, which is acceptable.  One of the main sources of difference is that in the numerical solution local stress fields and their relaxation by incomplete martensitic variants at the twin within twin interface is automatically taken into account.

\section{Concluding remarks}
\label{consc}
A thermodynamically consistent Ginzburg-Landau type nanoscale multiphase phase-field approach to multivariant martensitic transformations is formulated at large strain and taking the interfacial  stresses into consideration. A total of $N$ independent order parameters are assumed to describe the austenite and $N$ martensitic variants. A non-contradictory form for the gradient energy is considered, and the gradient energies with the previous models are compared. Furthermore,  a non-contradictory kinetic model for the coupled Ginzburg-Landau equations is derived for all the order parameters. The kinetic models from the previous models are compared and their shortcomings are discussed. A  general crystallographic solution for the twins within twins microstructure is obtained, and the solutions for cubic to tetragonal transformations are presented. Using the present phase-field approach 3D twins within twins microstructures evolution in a single grain is studied, and the numerical results are compared with the crystallographic solution. The present phase-field model can be used for studying more complex microstructures with more  variants such as cubic$\leftrightarrow$orthorhombic and cubic$\leftrightarrow$monoclinic MTs, and also in polycrystalline solids. Note that for large transformation strains for the MTs Si I to Si II, a new martensitic microstructure, which does not obey the mathematical theory of martensite  \cite{Adachi-Wayman-75,Bha04,Pitteri-Zanzotto-2003,Ball-James-87,Bhattacharya-1991}, was obtained with molecular dynamic simulations in \cite{Chen-Levitas-NC-22}. It will be a challenge for the current (and any other) large-strain theory to reproduce such a microstructure.

\noindent{\bf Acknowledgments}

AB acknowledges the support from SERB, Government of India (Project number SRG/2020/001194), and IIT Tirupati.
{VL acknowledges the support from NSF (CMMI-1943710 and DMR-1904830)    and Iowa State University (Vance Coffman Faculty Chair Professorship). The simulations were performed at  Extreme Science and
Engineering Discovery Environment (XSEDE), allocations TG-MSS140033 and MSS170015.}

\end{document}